\documentclass[aps,prx,reprint,showpacs,amsmath,amssymb,superscriptaddress,citeautoscript]{revtex4-2}
\usepackage{graphicx}
\usepackage{epstopdf}
\usepackage{color}
\usepackage{cmap}
\usepackage{graphicx}
\usepackage{epstopdf}
\usepackage{xcolor}
\usepackage{physics}
\usepackage{float}
\usepackage{amsmath}
\usepackage{bm}
\usepackage{hyperref}
\usepackage{setspace}

\begin{document}

\preprint{M. O. Ajeesh et al.}

\title{Interplay of structure and magnetism in LuFe$_4$Ge$_2$ tuned by hydrostatic pressure}

\author{M. O. Ajeesh} 
\email{Present address: Los Alamos National Laboratory, Los Alamos, New Mexico 87545, USA; ajeesh@lanl.gov}
\affiliation{Max Planck Institute for Chemical Physics of Solids, N\"{o}thnitzer Str.\ 40, 01187 Dresden, Germany}

\author{P. Materne} \affiliation{Advanced Photon Source, Argonne National Laboratory, Lemont, IL, 60439, USA}

\author{R. D. dos Reis} \affiliation{Brazilian Synchrotron Light Laboratory (LNLS), Brazilian Center for Research in Energy and Materials (CNPEM),
Campinas, Sao Paulo, Brazil}

\author{K. Weber} \affiliation{Max Planck Institute for Chemical Physics of Solids, N\"{o}thnitzer Str.\ 40, 01187 Dresden, Germany}

\author{S. Dengre} \affiliation{Institut für Festk\"{o}rper-und Materialphysik, Technische Universit\"{a}t Dresden, 01069 Dresden, Germany}

\author{R. Sarkar} \affiliation{Institut für Festk\"{o}rper-und Materialphysik, Technische Universit\"{a}t Dresden, 01069 Dresden, Germany}

\author{R. Khasanov} \affiliation{Laboratory for Muon Spin Spectroscopy, Paul Scherrer Institute, 5232 Villigen, Switzerland}

\author{I. Kraft} \affiliation{Max Planck Institute for Chemical Physics of Solids, N\"{o}thnitzer Str.\ 40, 01187 Dresden, Germany}

\author{A. M. Le\'{o}n} \affiliation{Max Planck Institute for Chemical Physics of Solids, N\"{o}thnitzer Str.\ 40, 01187 Dresden, Germany}

\author{W. Bi} \affiliation{Advanced Photon Source, Argonne National Laboratory, Lemont, IL, 60439, USA} \affiliation{Department of Physics, University of Alabama at Birmingham, Birmingham, AL 35294, USA}

\author{J. Zhao} \affiliation{Advanced Photon Source, Argonne National Laboratory, Lemont, IL, 60439, USA}

\author{E. E. Alp} \affiliation{Advanced Photon Source, Argonne National Laboratory, Lemont, IL, 60439, USA}

\author{S. Medvedev} \affiliation{Max Planck Institute for Chemical Physics of Solids, N\"{o}thnitzer Str.\ 40, 01187 Dresden, Germany}

\author{V. Ksenofontov} \affiliation{Institut f\"{u}r Anorganische und Analytische Chemie, Johannes Gutenberg-Universit\"{a}t, 55099 Mainz, Germany}

\author{H. Rosner} \affiliation{Max Planck Institute for Chemical Physics of Solids, N\"{o}thnitzer Str.\ 40, 01187 Dresden, Germany}

\author{H.-H. Klauss} \affiliation{Institut für Festk\"{o}rper-und Materialphysik, Technische Universit\"{a}t Dresden, 01069 Dresden, Germany}

\author{C. Geibel} \affiliation{Max Planck Institute for Chemical Physics of Solids, N\"{o}thnitzer Str.\ 40, 01187 Dresden, Germany}

\author{M. Nicklas} \email{Michael.Nicklas@cpfs.mpg.de} \affiliation{Max Planck Institute for Chemical Physics of Solids, N\"{o}thnitzer Str.\ 40, 01187 Dresden, Germany}

\date{\today}
\begin{abstract}

LuFe$_4$Ge$_2$ crystallizes in the ZrFe$_4$Si$_2$-type structure, hosting chains of Fe-tetrahedra giving rise to geometric frustration and low-dimensionality. The compound orders antiferromagnetically at around 36~K accompanied by a simultaneous structural transition from a tetragonal to an orthorhombic phase. The hydrostatic pressure dependence of the magnetic and structural transitions is investigated using electrical-transport, ac magnetic-susceptibility, ac calorimetry, M$\ddot{\rm o}$ssbauer, muon-spin relaxation ($\mu$SR), and x-ray diffraction measurements. External pressure suppresses the first-order transition to the antiferromagnetic phase (AFM1) around 1.8~GPa. The structural transition is largely unaffected by pressure and remains between 30 to 35~K for pressures up to 2~GPa. A second antiferromagnetic phase (AFM2) is observed at higher pressures. The transition from the paramagnetic to the AFM2 phase is of second-order nature and appears to be connected to the structural transition. The magnetic volume fraction obtained from $\mu$SR and M$\ddot{\rm o}$ssbauer measurements reveal that the entire sample undergoes magnetic ordering in both magnetic phases. In addition, similar low-temperature muon-precession frequencies in AFM1 and AFM2 phases point at similar ordered moments and magnetic structures in both phases. Our results further indicate enhanced magnetic fluctuations in the pressure induced AFM2 phase. The experimental observations together with density functional theory calculations suggest that the magnetic and structural order parameters in LuFe$_4$Ge$_2$ are linked by magnetic frustration, causing the simultaneous magneto-structural transition.

\end{abstract}

\maketitle

\section{INTRODUCTION}

Compounds with competing ground states have attracted tremendous attention in condensed matter research. This is because novel properties such as quantum criticality and unconventional superconductivity are often observed in regions of competing energy scales in a variety of material classes~\cite{Mathur98,Gegenwart08,Shibauchi14,Scalapino12, Keimer15, Gruner17}. In such compounds the interplay of magnetic, electronic, and structural degrees of freedom dictates the emerging phenomena. A prime example is the unconventional superconductivity observed in iron pnictides, where the role of magnetic and nematic fluctuations are crucially debated~\cite{Chubukov12,Dai15,Inosov16,Kou16,Wang22}. In this regard, compounds with low-dimensionality and magnetic frustration are of particular interest due to enhanced quantum fluctuations. Therefore identifying new systems and detailed investigations by tuning their ground-state properties are important for improving our understanding of unconventional phenomena.

\begin{figure}[t]
\centering
\includegraphics[width=1\linewidth]{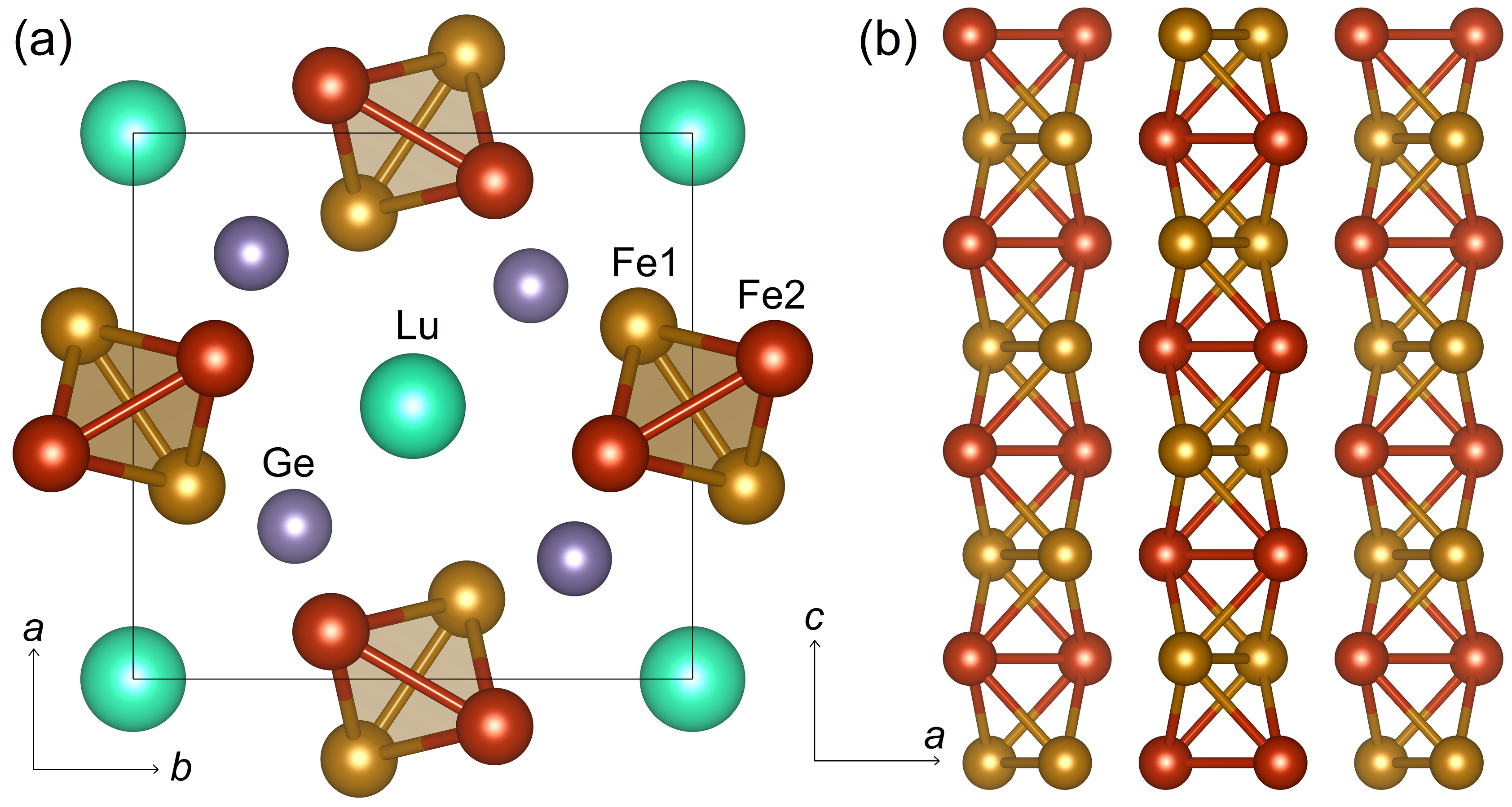}
\caption{(a) Crystal structure of LuFe$_4$Ge$_2$ viewed along the $c$ axis. In the low-temperature orthorhombic phase ($Pnnm$), the Fe sites split in to two sites marked as Fe1 and Fe2~\cite{Schobinger12}. (b) The chainlike arrangement of edge-shared Fe tetrahedra viewed along the $b$ axis.}
\label{Fig1}
\end{figure}

Intermetallic $A$Fe$_4X_2$ ($A=$ rare-earth, $X=$ Si, Ge) compounds are ideal candidates for studying unconventional phases and the effect of magnetic frustration on their properties. These compounds crystallize in the ZrFe$_4$Si$_2$-type structure (space group $P4_2/mnm$) consisting of a slightly distorted tetrahedral arrangement of Fe atoms, a geometry well-known for exhibiting magnetic frustration~\cite{Yarmoluk75}. Moreover, the Fe tetrahedra are edge-shared to form chains along the crystallographic $c$-axis, resulting in a quasi-one-dimensional structure (see Fig.~\ref{Fig1}). These quasi-one-dimensional chains of geometrically frustrated Fe tetrahedra form a very different type of magnetic lattice compared to the 122 Fe-pnictides where the magnetic frustration is caused by competing exchange interactions. In that way the intermetallic $A$Fe$_4X_2$ materials offer a new perspective on the entanglement of crystal structure and magnetism.

Previous investigations on the isostructural compound ZrFe$_4$Si$_2$ using chemical substitution and external pressure revealed that this compound is close to a lattice-volume tuned quantum critical point~\cite{Ajeesh20}. Furthermore, significantly large electronic heat capacity observed at low temperatures has been ascribed to the effect of magnetic frustration. The combination of magnetic frustration and low-dimensionality in these compounds draws the attention for a detailed investigation on other materials in the family.

Earlier studies on LuFe$_4$Ge$_2$ showed an antiferromagnetic (AFM) transition at 32~K with first-order character accompanied by a structural transition from tetragonal $P4_2/mnm$ to orthorhombic $Pnnm$~\cite{Schobinger12}. The results pointed at a canted arrangement of Fe moments in the $ab$-plane yielding a commensurate antiferromagnetic phase with propagation vector $q=0$. Moreover, the size of the ordered Fe moment of 0.44 $\mu_{\rm B}$ appeared to be highly reduced, which is attributed to the presence of magnetic frustration. However, a detailed study on the physical properties and the interplay of structure and magnetism is lacking.

In this article, we report a detailed investigation on the pressure evolution of the magnetic and structural transitions in LuFe$_4$Ge$_2$, carried out using electrical-transport, magnetic-susceptibility, ac calorimetry, M$\ddot{\rm o}$ssbauer spectroscopy, muon-spin relaxation, and x-ray diffraction experiments under external pressure. The nature of the phase transitions and the details of the various magnetic phases are discussed. Furthermore, our experimental findings together with theoretical calculations based on density functional theory elucidate the role of magnetic frustration in the interplay of magnetic and structural degrees of freedom in LuFe$_4$Ge$_2$. 

\section{RESULTS AND DISCUSSION}

\subsection{AMBIENT PRESSURE CHARACTERIZATION}

\begin{figure}[t]
\centering
\includegraphics[width=1\linewidth]{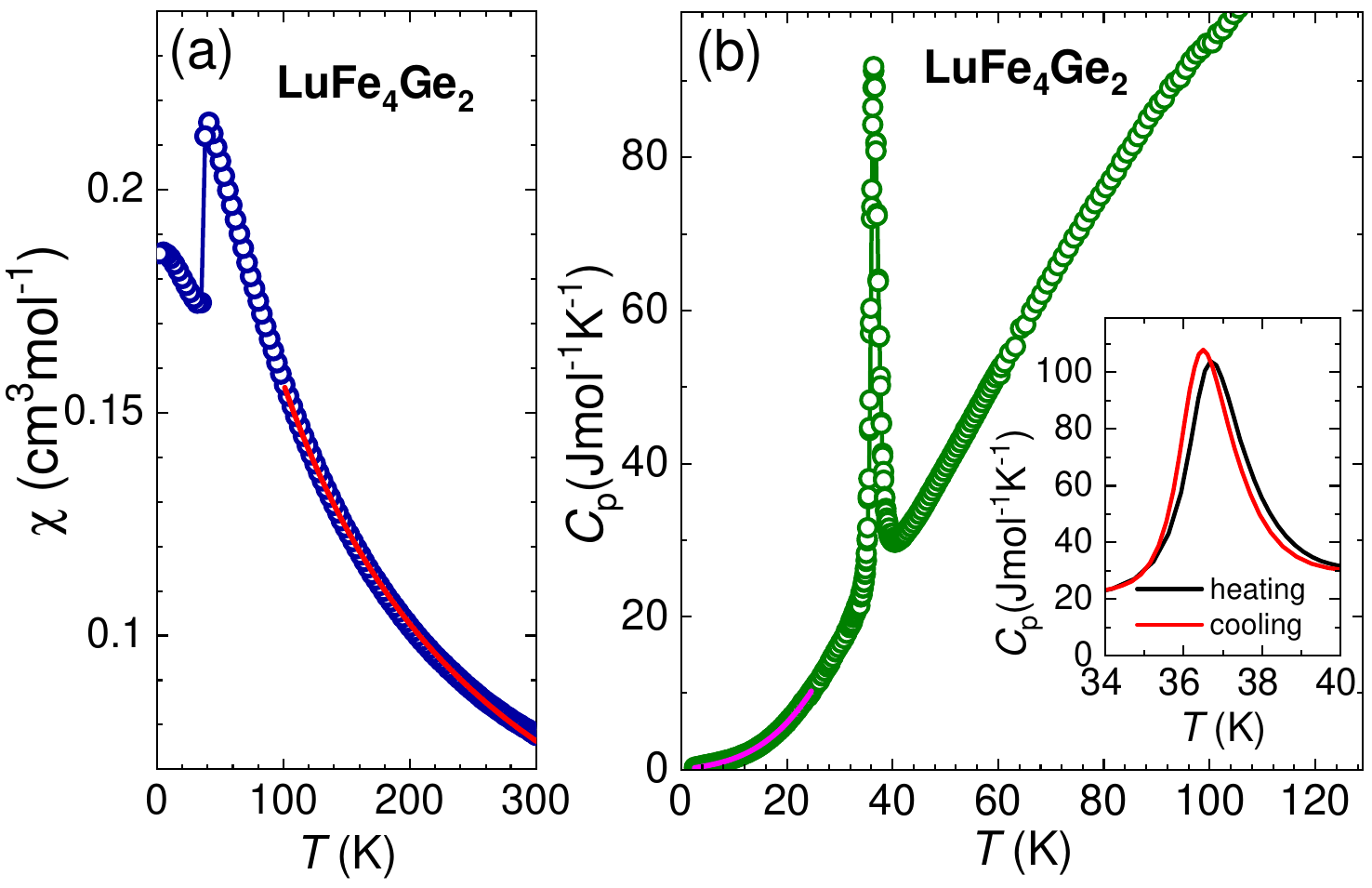}
\caption{(a) Temperature dependence of the dc magnetic susceptibility of LuFe$_4$Ge$_2$. The red curve is a Curie-Weiss fit to the data in
the temperature interval between 100 and 300~K. (b) Heat capacity $C_p$ of LuFe$_4$Ge$_2$ as a function of temperature. The line is a fit to the data for 2~K$\leq T \leq 25$~K using $C_p(T)=\gamma T+\beta T^3$. The inset displays the thermal hysteresis in $C_p(T)$ obtained from the analysis of heating and cooling parts of the thermal-relaxation cycle.}
\label{Fig2}
\end{figure}

The temperature dependence of the dc magnetic susceptibility $\chi=M/H$ of LuFe$_4$Ge$_2$ at ambient pressure is shown in Fig.~\ref{Fig2}a. Here, the ferromagnetic contribution in the magnetization from a small amount of Fe$_3$Ge phase ($<4$\%) in the sample is estimated by measuring magnetization at different fields, which is then subtracted from the measured data to obtain the intrinsic magnetic susceptibility of LuFe$_4$Ge$_2$. High-temperature $\chi(T)$ follows a Curie-Weiss (CW) behavior $\chi(T)=C/(T-\theta_{\rm W})$ with an effective moment $\mu_{\rm eff}$($=\sqrt{3K_{\rm B}C/N_{\rm A}\mu_0\mu_{\rm B}^2}$) of 2.2~$\mu_{\rm B}$/Fe and a Weiss temperature $\theta_{\rm W}=-90$~K. These CW parameters indicate relatively large Fe moments in the paramagnetic phase with dominant antiferromagnetic interaction among them. At low temperature, a sharp drop in susceptibility is observed at $T_N=36$~K corresponding to the antiferromagnetic transition. It should be noted that the observed $T_N$ differs from the earlier reported value of 32~K~\cite{Schobinger12}. This may be due to a difference in sample quality, where a large amount of impurity phases could have affected the precise determination of the ordering temperature using neutron diffraction studies.

Heat capacity $C_p$ of LuFe$_4$Ge$_2$ as a function of temperature is presented in Fig.~\ref{Fig2}b. A very sharp peak in $C_p(T)$ is observed at $T=36$~K, consistent with $T_N$ obtained from magnetic-susceptibility data. The low-temperature part of $C_p(T)$ is fitted with $C_p(T)=\gamma T+\beta T^3$ in the interval between 2 and 25~K. The fit yields an enhanced value for the Sommerfeld coefficient $\gamma=96$~mJ/molK$^2$ which indicates strong electron correlation effects. $C_p(T)$ upon heating and cooling, obtained by analyzing the thermal-relaxation curves recorded in a standard relaxation-type measurement setup following a method outlined by Lashley \textit{et al.\ }\cite{Lashley03}, is plotted in the inset of Fig.~\ref{Fig2}b. A thermal hysteresis observed between heating and cooling curves confirms the first-order nature of the magneto-structural transition.

\subsection{EXPERIMENTS UNDER HYDROSTATIC PRESSURE}
\subsubsection*{Electrial Resistivity}

\begin{figure}[tb]
\centering
\includegraphics[width=1\linewidth]{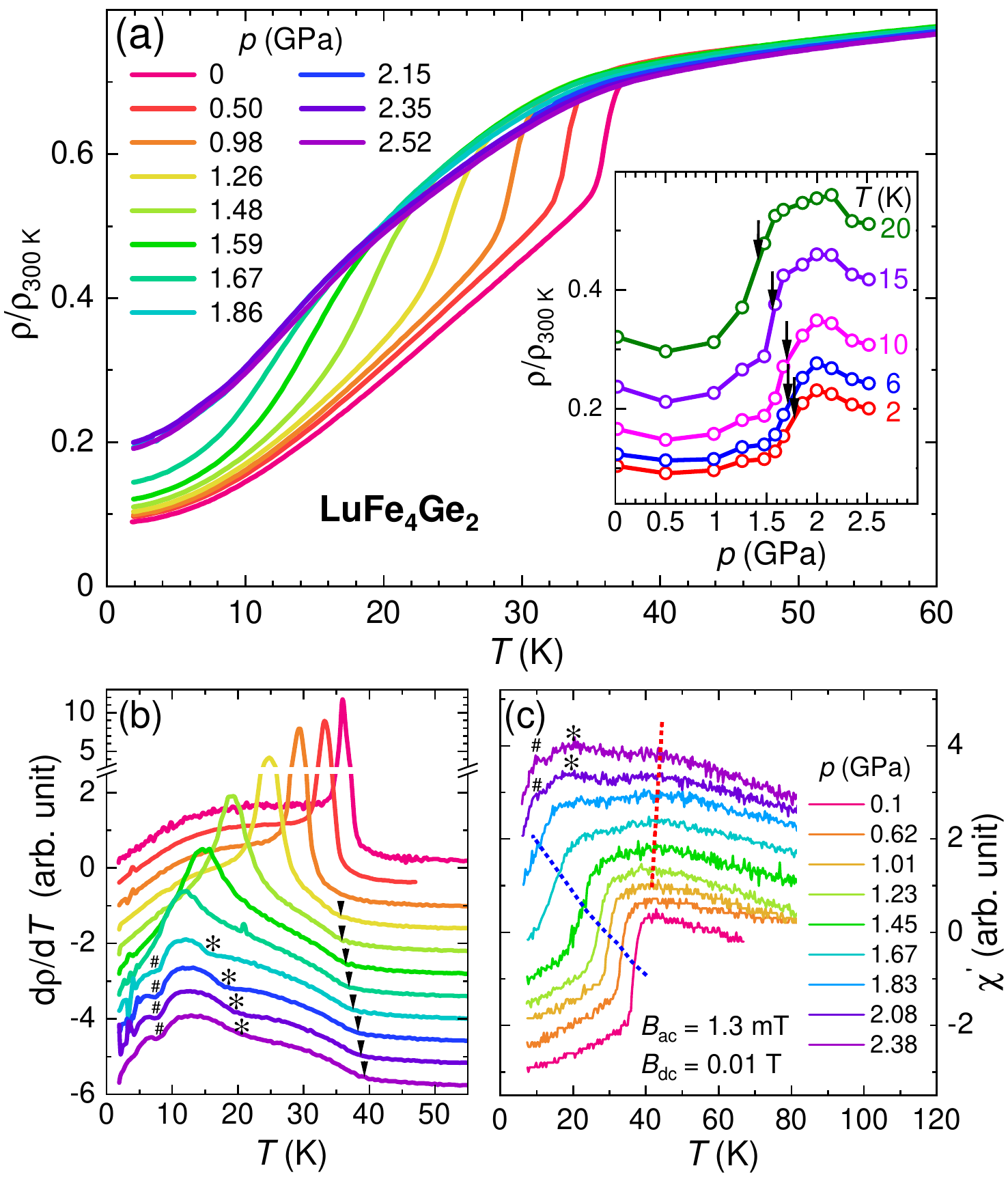}
\caption{(a) Electrical resistivity of LuFe$_4$Ge$_2$ as a function of temperature for several applied pressures. Inset: Resistivity isotherms as a function of pressure. The arrows indicate the phase boundary to a pressure-induced phase. (b) Temperature derivative of electrical resistivity $d\rho/dT$ vs.\ T for several pressures. Various anomalies are marked by symbols. (c) Temperature dependence of the real part of the ac magnetic susceptibility ($\chi'$) of LuFe$_4$Ge$_2$ for selected applied pressures. The pressure dependence of the anomalies are traced by the dashed lines and additional low-temperature anomalies at higher pressures are marked by $\ast$ and \# symbols}
\label{Fig3}
\end{figure}

To investigate the evolution of the magneto-structural transition upon application of hydrostatic pressure we first turn to electrical-resistivity measurements. Figure~\ref{Fig3}a presents the evolution of the temperature dependence of the electrical resistivity $\rho(T)$ under external pressure. At ambient pressure, LuFe$_4$Ge$_2$ shows metallic behavior upon cooling followed by a sudden drop in the resistivity at 36~K coinciding with the magneto-structural transition. The decrease in the resistivity at the transition temperature can be understood as the reduction in the scattering contribution due to the transition from the disordered paramagnetic to the antiferromagnetic phase. Upon further cooling, the resistivity decreases with a higher slope, indicating a further reduction of the scattering in the antiferromagnetic phase. The resistivity ratio (RR) at ambient pressure is determined as $\rho_{\rm 300K}/\rho_{\rm 1.8K}\approx11$ for the investigated sample. Under pressure, initially, the anomaly in the resistivity shifts to lower temperatures with increasing pressure while the sharp nature of the anomaly changes to a gradual decrease. The anomaly becomes much weaker and moves to around 12~K at a pressure of 1.7~GPa. Above 1.7~GPa, the feature becomes too small and not traceable within the resolution of our data. However, the resistivity isotherms plotted as a function of pressure (see the inset of Fig.~\ref{Fig3}a) display clear jumps at the phase boundary. These data suggest that the first-order-like antiferromagnetic transition (AFM1) is suppressed to zero temperature by the application of a pressure of $p_c\approx1.8$~GPa. In addition, a weak shoulder-like anomaly in $\rho(T)$ develops around $T=35$~K which becomes more prominent at higher pressures. This anomaly is better visible in the temperature derivative of the electrical resistivity $d\rho/dT$ presented in Fig.~\ref{Fig3}b. With increasing pressure this feature slowly shifts to higher temperatures, reaching about 40~K at $p=2.52$~GPa, suggesting a phase boundary from the PM phase to a pressure-induced phase. Furthermore, for $p\geq 1.86$~GPa, two additional features are clearly seen in $d\rho/dT$ at low temperatures around 20~K and 10~K (marked by $\ast$ and \# symbols, respectively). We note that the residual resistance above $p_c$, in the  pressure-induced phase, is considerably larger than that in the AFM1 phase at low pressures.

\subsubsection*{Magnetic Susceptibility}

The results from $\rho(T)$ measurements are corroborated by ac magnetic susceptibility which further confirm their magnetic origin. The results of the ac magnetic-susceptibility measurements on LuFe$_4$Ge$_2$ for several applied pressures are presented in Fig.~\ref{Fig3}c. The susceptibility data at each pressure were normalized by the jump height of the superconducting transition of the Pb manometer in order to compare the data taken at different pressures. At $p=0.1$~GPa, the real part of ac susceptibility ($\chi'$) shows a sudden decrease at the antiferromagnetic transition. Upon increasing pressure the drop in $\chi'$ shifts to lower temperatures while the sharpness of the jump is reduced. A broad humplike feature develops around 40~K and shows a slight shift to higher temperatures with further increase in pressure. These observations are in excellent agreement with the resistivity data. Above 1.8~GPa, the nature of the $\chi'(T)$ curve is significantly different from that in the low pressure region. In the high pressure region ($p>1.8$~GPa), an additional humplike feature and a sharp decrease in the susceptibility are observed at about 20~K and 10~K coinciding with the low-temperature anomalies observed in $d\rho/dT$. The origin and the nature of these anomalies are not fully understood, however, their implication in other physical properties are discussed later.

\begin{figure}[tb]
\centering
\includegraphics[width=1\linewidth]{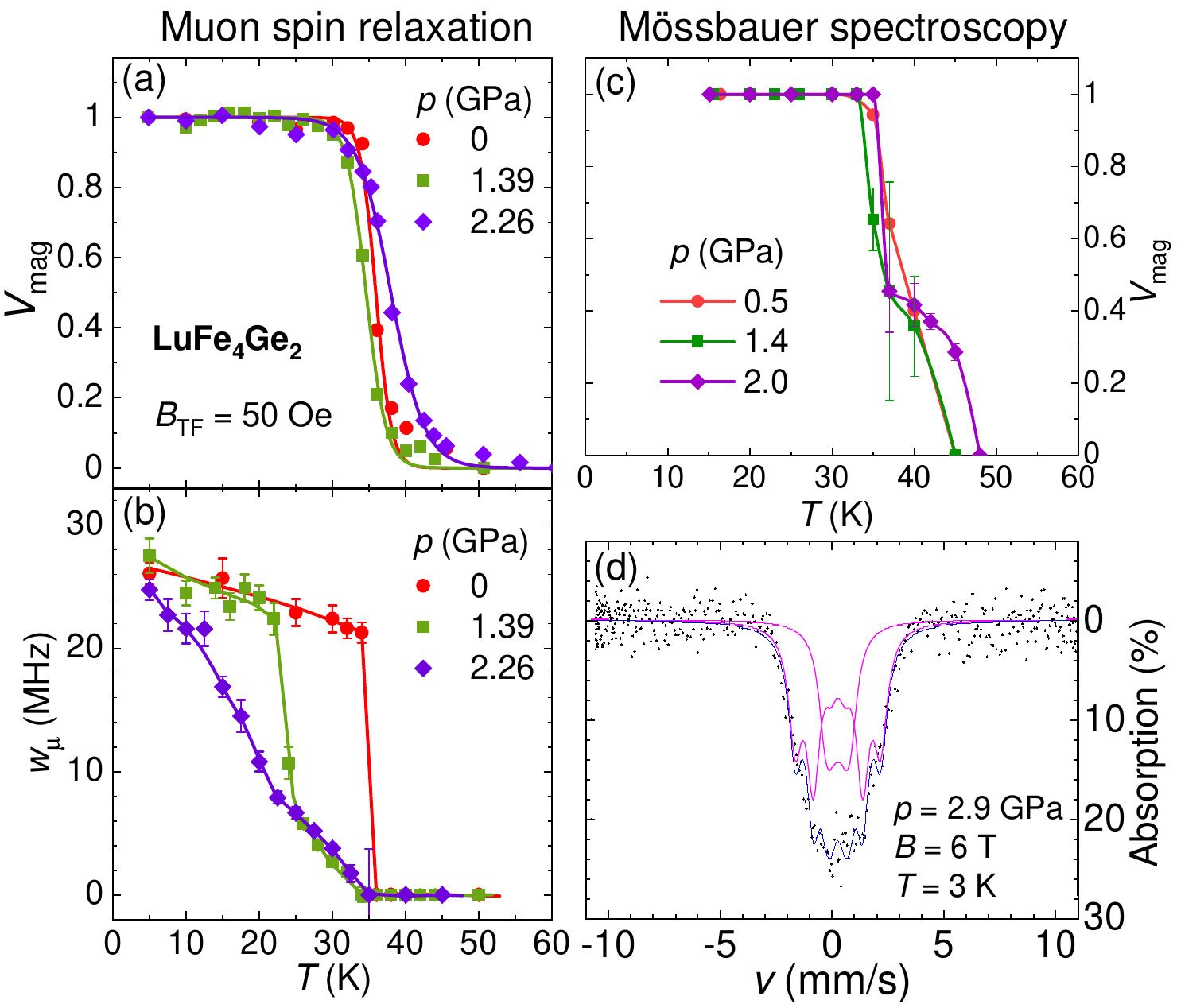}
\caption{(a) Temperature dependence of the magnetic-volume fraction $V_{\rm mag}$ of LuFe$_4$Ge$_2$ obtained using $\mu$SR at different pressures. The solid lines are fits to the data using a phenomenological equation provided in the text. (b) Temperature dependence of the muon-spin precession frequency under different applied pressures. (c) $V_{\rm mag}$ obtained from time-domain $^{57}$Fe M$\ddot{\rm o}$ssbauer spectroscopy under several pressures. (b) Energy-domain $^{57}$Fe M$\ddot{\rm o}$ssbauer spectra measured at $p=2.9$~GPa, $T=3$~K, and an external magnetic field of 6~T. The solid lines are fits to the spectra by \emph{Exact Lineshape Site Analysis} using RECOIL software.}
\label{Fig4}
\end{figure}

\subsubsection*{M$\ddot{\rm o}$ssbauer and Muon-Spin Spectroscopy}

In order to get a microscopic understanding of the nature of the different magnetic phases of LuFe$_4$Ge$_2$, we have carried out muon-spin relaxation ($\mu$SR) and $^{57}$Fe M$\ddot{\rm o}$ssbauer spectroscopy measurements under several pressures below and above $p_c\approx1.8$~GPa. $\mu$SR measurements in a weak transverse-field of $B_{\rm TF}=50$~Oe were performed to determine the magnetic-volume fraction ($V_{\rm mag}$) and the transition temperatures. In Fig.~\ref{Fig4}a, $V_{\rm mag}$ of LuFe$_4$Ge$_2$ as a function of temperature for three different pressures is shown. $V_{\rm mag}(T)$ shows a sharp step-like increase at the magnetic ordering temperature for all pressures, a characteristic feature of long-range magnetic order. It is also important to note that the entire sample volume undergoes magnetic ordering at all pressures. The data confirm that the pressure-induced phase is also long-range magnetically ordered. The transition temperature $T_N$ is extracted by fitting the magnetic-volume fraction with the phenomenological equation $V_{\rm mag} = \frac{1}{1+e^{(T-T_N)/w}}$ where $w$ is the transition width. Here, the magnetic ordering temperature seems to be nearly independent of pressure and coincide with the weak $p$-independent anomalies in the resistivity and susceptibility data. A 100\% magnetic volume fraction in both phases is also confirmed by M$\ddot{\rm o}$ssbauer data (see Fig.~\ref{Fig4}c). Furthermore, the M$\ddot{\rm o}$ssbauer spectrum obtained at $p=2.9$~GPa, $T=3$~K, and an external magnetic field of 6~T (see Fig.~\ref{Fig4}d) suggests that the pressure-induced phase is antiferromagnetically ordered (AFM2). The fit to the spectrum by \emph{Exact Lineshape Site Analysis} using RECOIL software revealed contributions from Fe moments parallel and antiparallel to the external magnetic field. The fraction of magnetic moments on the Fe parallel to external field is about 66\% with a hyperfine magnetic field of $B_{\rm hf} = 11.8$~T. This value is higher than that of the 34\% of magnetic moments on Fe antiparallel to external field with $B_{\rm hf} = 3.6$~T. This is presumably because of the rotation of some part of the moments to the direction of the applied magnetic field. Such an observation is consistent with the metamagnetic transition seen in the magnetoresistance data (see Appendix). In the magnetoresistance (MR) data, the spin reorientation appears to occur at external fields as low as 5~T at high pressures.

\begin{figure}[t]
\centering
\includegraphics[width=1\linewidth]{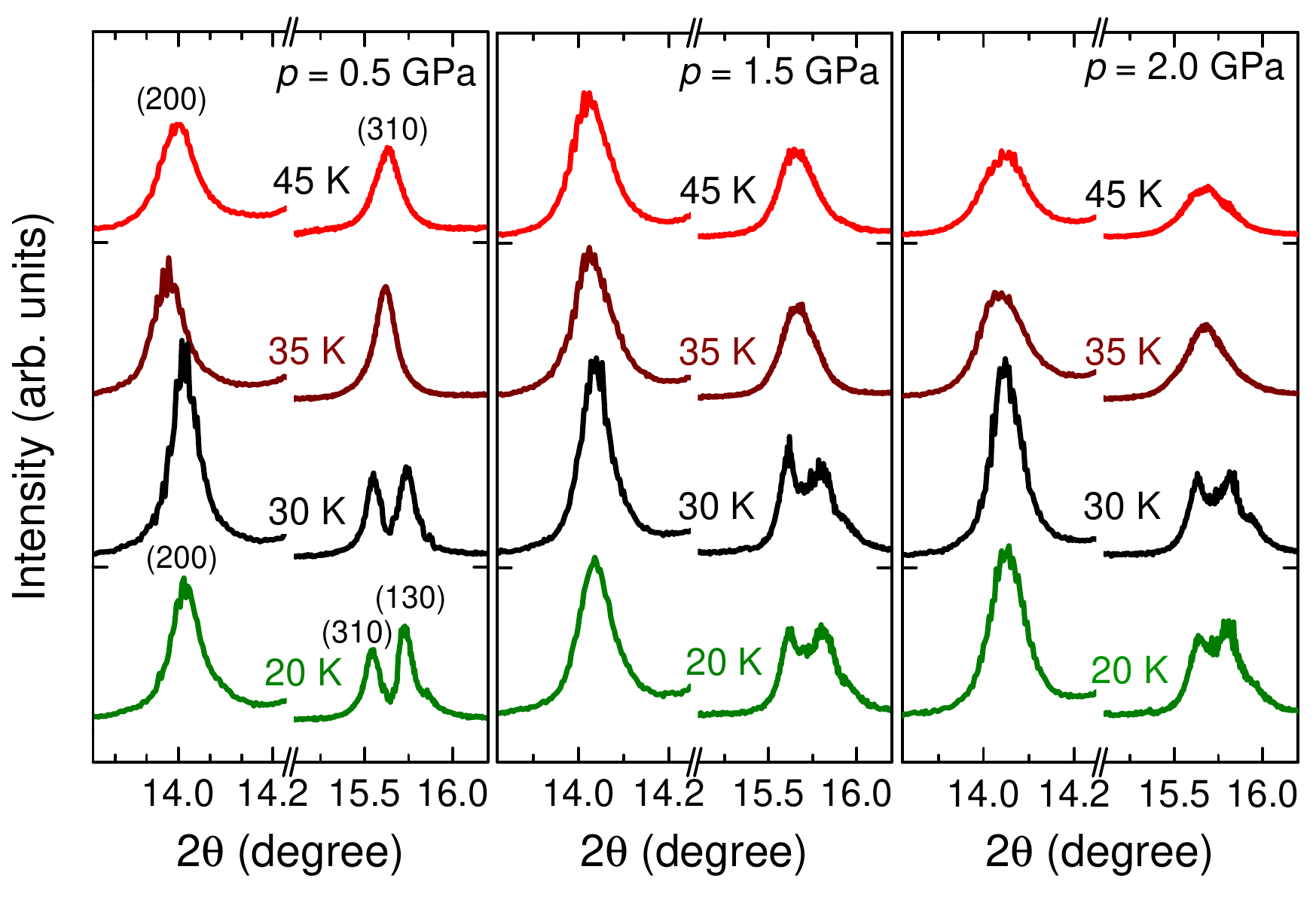}
\caption{Representative peaks in the X-ray diffraction patterns of LuFe$_4$Ge$_2$ at different temperatures and applied pressures. The structural transition from the tetragonal to the orthorhombic phase is evidenced by the splitting of the diffraction peak around $2\theta\approx15.5^0$.}
\label{Fig5}
\end{figure}

Further information regarding the strength of the local magnetic field $B_{\rm loc}$ at the muon site in the magnetically ordered phases is obtained from zero-field (ZF) $\mu$SR measurements. The temperature dependence of the muon-precession frequency $\omega_{\mu}$ for three different pressures is displayed in Fig.~\ref{Fig4}b. At ambient pressure, the spontaneous precession occurs at $T\approx36$~K with a slight enhancement in $\omega_{\mu}$ upon further decreasing temperature. This sharp step-like behavior of $\omega_{\mu}(T)$ is consistent with the first-order nature of the phase transition at ambient pressure. At $p=1.39$~GPa, the precession starts at $T\approx34$~K. The frequency gradually increases upon cooling followed by a sudden jump at $T\approx26$~K. These features can be well ascribed to the two consecutive phase transitions upon lowering temperature: the first from the paramagnetic to the pressure-induced AFM2  phase at 34~K and the second from the AFM2 to  the AFM1 phase at 26~K. The gradual increase in $\omega_{\mu}$ at the first phase transition implies that this transition is of second order-type. The sharp jump in $\omega_{\mu}$ around 26~K arises from the transition from AFM2 to AFM1 phase and the sharpness  points at a first-order-type phase transition. At $p=2.26$~GPa, the precession starts around 35~K, where the compound undergoes the transition from the PM to the AFM2 phase. Upon further cooling, $\omega_{\mu}$ increases gradually until the lowest temperature in our experiment. We note that the $\omega_{\mu}(T)$ curve for $p=2.26$~GPa shows noticeable slope changes at temperatures around 20~K and 10~K. These features occur at the same temperatures as the low-$T$ anomalies in the $d\rho(T)/dT$ and $\chi'(T)$ data discussed earlier and suggest the possibility of multiple order-parameters associated with the phase transition from the paramagnetic to the high pressure AFM2 phase. It is worth noting that, at the lowest temperature, the muon-precession frequencies for the three pressures have similar values. This indicates that, at low temperature, both the AFM1 and the pressure-induced AFM2 phase have similar local magnetic fields at the muon sites.

\subsubsection*{X-ray Diffraction}

We now turn to the evolution of the structural transition in LuFe$_4$Ge$_2$ under external pressure. The structural transition from tetragonal to orthorhombic symmetry is characterized by the splitting of certain peaks in the diffraction pattern. For example, the  $[310]$  peak splits in the orthorhombic phase while the $[200]$ peak does not. Therefore, the evolution of these peaks with varying conditions of temperature and pressure provides a straight forward determination of the structural transition in the phase diagram. Figure~\ref{Fig5} presents the $[200]$  and $[310]$ diffraction peaks obtained at various temperatures and pressures. Remarkably, the splitting of the $[310]$ peak occurs between 30 and 35~K in the entire pressure range. This confirms that the structural transition temperature does not significantly change with pressure and it remains between 30 and 35~K. The lack of a more precise determination of the structural transition temperature is due to the limited temperature sampling available in our experiment.

\subsection{PHASE DIAGRAM}

\begin{figure}[t]
\centering
\includegraphics[width=1\linewidth]{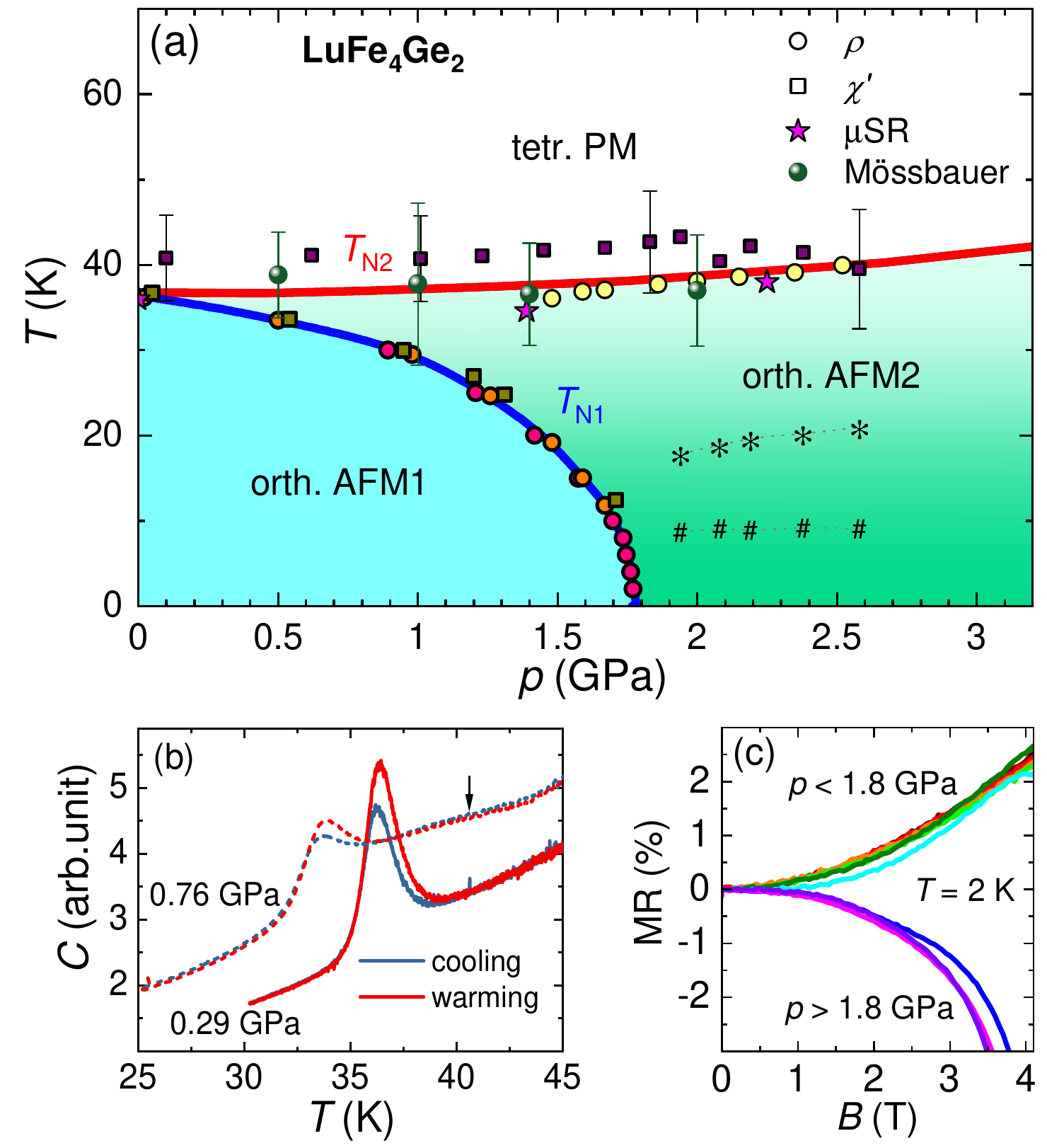}
\caption{(a) $T-p$ phase diagram of LuFe$_4$Ge$_2$. The transition temperatures obtained from $\rho(T)$ (circle), $\chi'(T)$ (square), $\mu$SR (star), and  M$\ddot{\rm o}$ssbauer (sphere) data are indicated. The $\ast$ and \# symbols stand for the low-temperature anomalies observed in $\rho(T)$ and $\chi'(T)$ data. (b) Thermal hysteresis in $C(T)$, measured using an ac-calorimetry technique under pressure, confirms the first-order nature of the transition at $T_{N1}$. The arrow points at a very weak feature at $T_{N2}$ (c) Magnetoresistance ${\rm MR}(B)=[\rho(B)-\rho(B=0)]/\rho(B=0)$ of LuFe$_4$Ge$_2$ measured at $T=2$~K.}
\label{Fig6}
\end{figure}

The temperature--pressure phase diagram of LuFe$_4$Ge$_2$ is established using the results from the different high-pressure studies, see Fig.~\ref{Fig6}a. The results from bulk measurements as well as from microscopic probes are in excellent agreement. $T_{N1}$ corresponding to the transition from the PM to the AFM1 phase is continuously suppressed toward zero temperature upon increasing pressure at a critical pressure of $p_c\approx1.8$~GPa. A second antiferromagnetic phase (AFM2) is confirmed by M$\ddot{\rm o}$ssbauer and $\mu$SR measurements. The structural transition temperature which coincide with the AFM1 ordering temperature at ambient pressures does not significantly change with application of pressure and remains at about 35~K, i.e. the AFM1 transition decouples from the structural transition. Moreover, the onset of the magnetic ordering from the paramagnetic (PM) to the AFM2 phase appears to be connected to the structural transition, i.e. its transition temperature $T_{N2}$ is also almost independent of pressure.

The nature of the transitions between the various phases have been studied in detail. As described earlier, the magneto-structural transition at ambient pressure is of first-order type. The application of pressure decouples the $T_{N1}$ and the structural transition. Figure~\ref{Fig6}b depicts heat-capacity data (measured using an ac-calorimetry technique) taken upon heating and cooling cycles under selected pressures of 0.29 and 0.76~GPa. The peak in $C(T)$ associated with the transition from  the AFM2 to the AFM1 phase shows a thermal hysteresis, confirming the first-order character of the transition. We note that the thermal hysteresis was measured with a very slow temperature sweep-rate in order to minimize the effect of the relatively slow thermalization of the pressure cell. A similar hysteresis is also observed in resistivity and ac magnetic-susceptibility data (not shown). The first-order nature of the transition is also consistent with the sharp jump in $\omega_{\mu}$ at this phase boundary. The transition from the PM to the AFM2 phase appears as very weak features in $C(T)$, $\rho(T)$, and $\chi'(T)$ without any observable thermal hysteresis. Moreover, $\omega_{\mu}$ shows a gradual increase at this phase boundary, confirming that the transition from the PM to the AFM2 phase is of second order. 

The magnetoresistance and the magnetic-susceptibility data provide insights into the relative strength of magnetic fluctuations in the two antiferromagnetic phases of LuFe$_4$Ge$_2$. In the AFM1 phase, the ${\rm MR}=[\rho(B)-\rho(B=0)]/\rho(B=0)$ exhibits a monotonous increase with a quadratic dependence on the magnetic field (see Fig.~\ref{Fig6}c). This is the typical behavior expected for metallic systems and is due to the cyclotron motion of the conduction electrons in the transverse magnetic field. However, in the AFM2 phase, ${\rm MR}(B)$ shows an opposite behavior displaying negative values. These observations can be taken as indication for stronger magnetic fluctuations in the AFM2 phase compared to the AFM1 phase. The negative MR stems from the reduction in the resistivity as the external magnetic field suppresses the magnetic fluctuations. The small reduction in the $\chi'$ at the magnetic ordering at the phase boundary to AFM2 phase compared to the large drop in $\chi'$ at the ordering at the phase boundary to AFM1 phase is also in support of the existence of strong magnetic fluctuations in AFM2 phase. Moreover, the gradual increase in the muon-precession frequency, which is a measure of the local magnetic field, in the AFM2 phase upon deceasing temperature suggests a strong temperature dependence of the magnetic fluctuations.

\begin{figure}[t]
\centering
\includegraphics[width=1\linewidth]{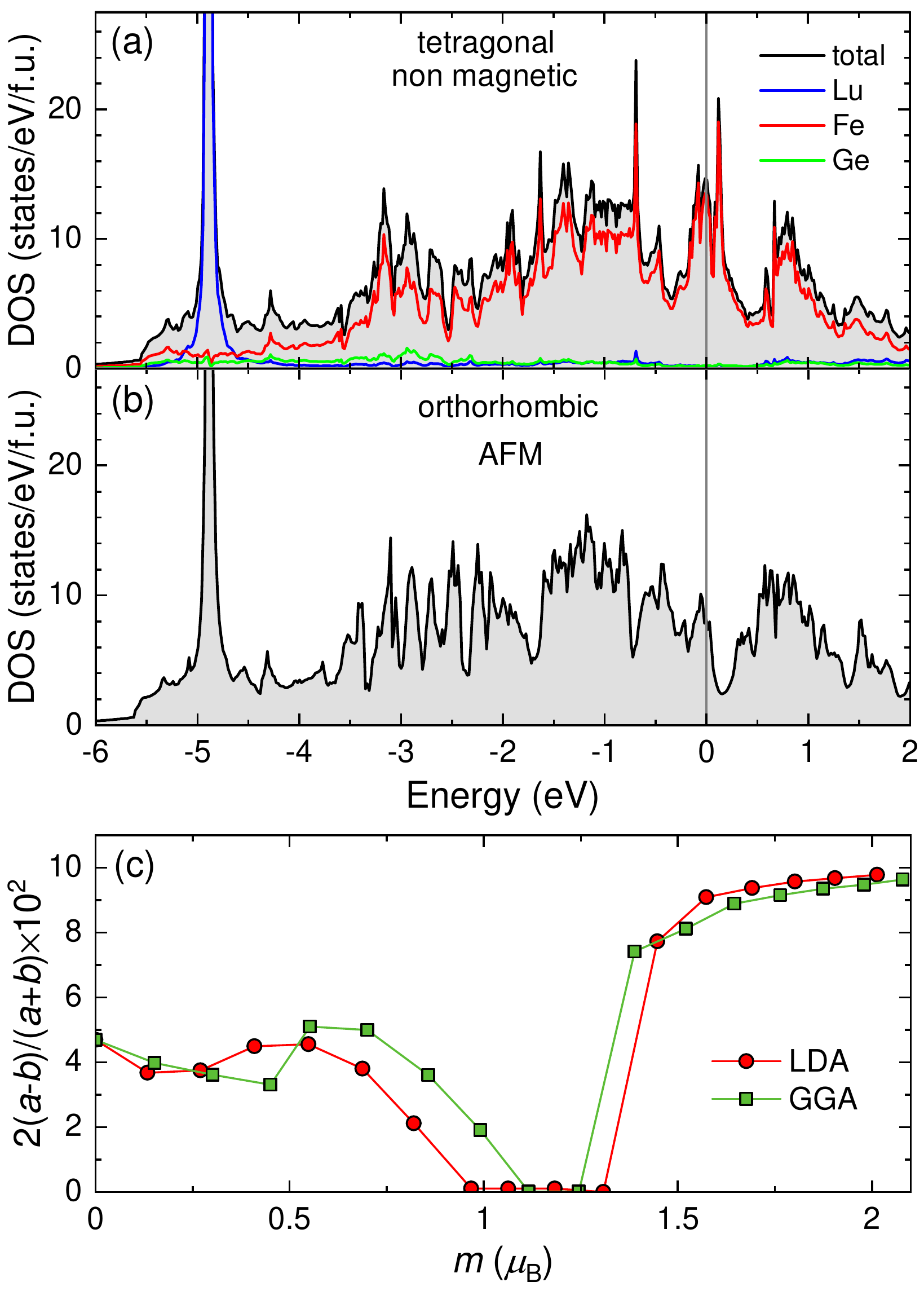}
\caption{Total and partial density of states of LuFe$_4$Ge$_2$ for (a) the non-magnetic tetragonal and (b) the antiferromagnetic orthorhombic case. The Fermi level is at zero energy.  Panel (c) shows the dependence of the lattice distortion (in \%) on the Fe magnetic moment for different exchange correlation potentials.}
\label{Fig7}
\end{figure}

\subsection{ELECTRONIC STRUCTURE CALCULATIONS}

Electronic-structure calculations for LuFe$_4$Ge$_2$ based on density functional theory (DFT) have been performed to understand its magnetic and structural phase transitions. The magnetic properties are determined by the dominating contribution of the Fe $3d$ related rather narrow bands near the Fermi level $E_F$, resulting in a pronounced peak in the density of states (DOS) near $E_F$ (see Fig.~\ref{Fig7}a). With a value of about 3.5 states (per Fe and eV) at the Fermi level, the Stoner criterion is fulfilled, evidencing a magnetic instability and the tendency for spontaneous magnetic ordering. The magnetic and structural transition to a collinear AFM state leads to a strong reduction of the DOS at $E_F$ (compare Fig.~\ref{Fig7} a and b). To a large part, this drop in the DOS($E_F$) is caused by the band splitting related to the localized Fe $3d$ moments and thus similar for different magnetic structures.

A full relaxation of the crystal structure (at the experimental low-temperature volume) yields that the orthorhombic symmetry is energetically more stable than the tetragonal. Interestingly, the resulting equilibrium distortion $2(a-b)/(a+b)$ of the lattice is strongly dependent on the size of the magnetic moment, essentially independent from the choice of the exchange correlation potential (see Fig.~\ref{Fig7}c). This strong and non-linear dependence indicates that the magnetism and the crystal lattice are coupled in a nontrivial way. The energy dependence of the DOS for both the  tetragonal and the orthorhombic experimental crystal structures  {\it without} spin polarization reveals that the magnetic ordering can occur independent of the structural transition since for both symmetries we obtain a similarly large DOS near the Fermi level. Therefore, the reason for the magnetic ordering occurring simultaneously with the structural transition might not be directly related to the electronic band structure but rather due to the change in the strength of the magnetic frustration due to the lattice distortion. High-field magnetization measurements on LuFe$_4$Ge$_2$ showed that the magnetization remains small up to high fields and the extrapolated saturation field is more than 150~T~\cite{Weber17}. This corresponds to strong antiferromagnetic exchange interactions on the scale of more than 100~K. The strong exchange interaction along with a relatively low ordering temperature points at the significance of the magnetic frustration. It is likely that the magnetic frustration, at least to some extent, is released by the distortion of the Fe tetrahedra during the structural transition from the tetragonal to the orthorhombic phase and, thereby, facilitating the magnetic ordering. This scenario is supported by the calculations which predict the structural transition even without involving magnetic polarization. Here the large magnetic entropy connected with a fluctuating frustrated paramagnetic system, which stabilizes the tetragonal phase upon increasing temperature, might explain why the tetragonal phase is stable down to such comparatively low temperature. Our results suggest a mechanism where the magnetic and structural order parameters in LuFe$_4$Ge$_2$ are linked by the magnetic frustration causing the simultaneous magneto-structural transition. 

The commensurate magnetic ground state proposed by Papamantellos \emph{et al.}~\cite{Schobinger12} is also reproduced by our calculations. Moreover, a collinear spin structure is found to be energetically close to the non-collinear ground state. This points at the possibility that the pressure-induced AFM2 phase could be a spin rearrangement from non-collinear to collinear structure. Such a subtle spin rearrangement is also consistent with the similar values of low-temperature muon-precession frequency found in AFM1 and AFM2 phases.

\section{CONCLUSION}

In summary, LuFe$_4$Ge$_2$ presents an interesting interplay of magnetic, structural, and electronic degrees of freedoms. At ambient pressure LuFe$_4$Ge$_2$ undergoes a simultaneous antiferromagnetic and structural transition at 36~K with first-order character. The pressure dependence of the magnetic transition in LuFe$_4$Ge$_2$ has been investigated using electrical-transport, ac magnetic-susceptibility, ac calorimetry, M$\ddot{\rm o}$ssbauer, $\mu$SR, and PXRD measurements. External pressure suppresses the first-order magnetic transition (AFM1) to zero temperature around 1.8~GPa. The structural transition is largely unaffected by pressure. A new antiferromagnetic phase (AFM2) is observed at higher pressures, confirmed by M$\ddot{\rm o}$ssbauer and $\mu$SR measurements. The transition from the paramagnetic to the AFM2 phase is of second order and appears to be connected to the structural transition. $\mu$SR and M$\ddot{\rm o}$ssbauer data revealed 100\% magnetic volume fraction in both the magnetic phases. In addition, similar values of muon-precession frequency at low temperatures in the AFM1 and AFM2 phases point at similar ordered moments and closely related magnetic structures in the two phases. Our results also indicate enhanced magnetic fluctuations in the pressure-induced AFM2 phase. The experimental observations are supported by DFT band-structure calculations, suggesting a scenario where the magnetic and structural order parameters in LuFe$_4$Ge$_2$ are linked by magnetic frustration, causing the simultaneous magneto-structural transition.
Our results reveal an interesting and unusual interplay of structure and magnetism in LuFe$_4$Ge$_2$, which differ from the situation observed in the {\it A}Fe$_2$As$_2$ pnictides. Therefore, LuFe$_4$Ge$_2$ is an attractive system where further in-depth studies could provide deeper insight into the interaction between frustrated magnetism and structural instability, a topic of general interest which is also relevant for other classes of quantum materials.

\section{METHODS}

Polycrystalline samples of LuFe$_4$Ge$_2$ were synthesized by a standard arc-melting technique on a copper hearth. Constituent elements (at least 99.9\% purity), taken in the stoichiometric ratio, were melted in an arc furnace under argon atmosphere, followed by several flipping and remelting of the resulting ingot to ensure homogeneity. Then the as-cast samples were annealed at 1150$^0$C under a static argon atmosphere for a week. The phase purity of the annealed samples was checked by powder x-ray diffraction (PXRD) using Cu K$\alpha$ radiation and a scanning electron micrograph (SEM), which revealed only a small amount (up to 4\%) of eutectic phase Fe$_3$Ge in our samples. The stoichiometry of the samples was verified using energy dispersive x-ray (EDX) analysis.

DC magnetic susceptibility was measured using a superconducting quantum interference device (SQUID) magnetometer (magnetic properties measurement system, Quantum Design). The heat capacity at ambient pressure was recorded by a thermal-relaxation method using a physical property measurement system (PPMS, Quantum Design). Electrical-transport, ac magnetic-susceptibility, and ac-calorimetry measurements under hydrostatic pressure were performed using a double-layered piston-cylinder-type pressure cell with silicon oil as the pressure-transmitting medium. The pressure inside the sample space was determined at low temperatures by the shift of the superconducting transition temperature of a piece of Pb. The electrical resistivity was measured using a standard four-terminal method, where electrical contacts to the sample were made using 25~$\mu$m gold wires and silver paint. Resistivity was measured using an LR700 resistance bridge (Linear Research). AC magnetic susceptibility was measured using home-made ac-susceptometer that fits inside the pressure cell. The signal was measured using an LR700 mutual-inductance bridge (Linear Research). A static field $B_{\rm dc}$ of 0.01~T and a modulation field $B_{\rm ac}$ of 1.3~mT at a frequency of 16~Hz were used for the measurements.
AC calorimetry measurements were performed using a commercial heater and Cernox thermometer following the method described in Ref.~\cite{Gati19}.

Muon-spin relaxation measurements under pressure were performed at the $\mu$E1 beam-line using GPD spectrometer at the Paul Scherrer Institute (PSI), Switzerland (see Ref.~\cite{Khasanov16, Khasanov22} for details of the high-pressure $\mu$SR technique at PSI). Synchrotron $^{57}$Fe Mössbauer spectroscopy measurements under pressure were conducted at the Nuclear Resonance beamline (ID18) of the European Synchrotron Radiation Facility (ESRF), Grenoble~\cite{Potapkin12} and at the beamline 3ID-B of the Advanced Photon Source, Argonne National Laboratory, USA~\cite{Bi15, Zhao17}. X-ray diffraction experiments using a diamond-anvil cell for generating pressure were performed at the XDS beamline at the Brazilian Synchrotron facility LNLS~\cite{Lima16}.

DFT calculations were performed using the plane-wave pseudopotential method implemented in the Vienna ab-initio simulation package (VASP) \cite{VASP}, applying the local density approximation (LDA) \cite{PW} and the general gradient approximation (GGA) \cite{PBE} for the exchange-correlation functional. We use a plane-wave energy cutoff of 500 eV and a Regular Monkhorst-Pack grid of $8\times8\times12$ to perform the ionic relaxation and $10\times10\times14$ to achieve the self-consistent calculations. To obtain the density of states, the $k$-mesh was increased to $22\times22\times24$ using the tetrahedron method. The optimization of the structures was carried out with a force convergence tolerance of 1~meV/\AA\ per atom. The tetragonal (T) and orthorhombic (O) structures have $P4_{2}/mnm$ and $Pnnm$ space group symmetry, respectively. In order to study the FM and various AFM magnetic states and respective structural distortion of the orthorhombic structure, we carried out collinear and non-collinear calculations. To perform our studies, we have considered the structural parameters given in this study at 10~K and ambient pressure and the structural data given in the previous work \cite{Schobinger12}.

\section*{ACKNOWLEDGMENTS} 
This work was partly supported by Deutsche Forschungsgemeinschaft (DFG) through Research Training Group GRK 1621. U. Nitzsche is acknowledged for technical support for DFT calculations. RDdR acknowledges financial support from the Sao Paulo Research Foundation (FAPESP) (Grant 2018/00823-0) and from the Max Planck Society under the auspices of the Max Planck Partner Group R. D. dos Reis of the MPI for Chemical  Physics of Solids, Dresden, Germany.

\section*{APPENDIX: Field-induced metamagnetic transition}

\begin{figure}[t!]
\centering
\includegraphics[width=1\linewidth]{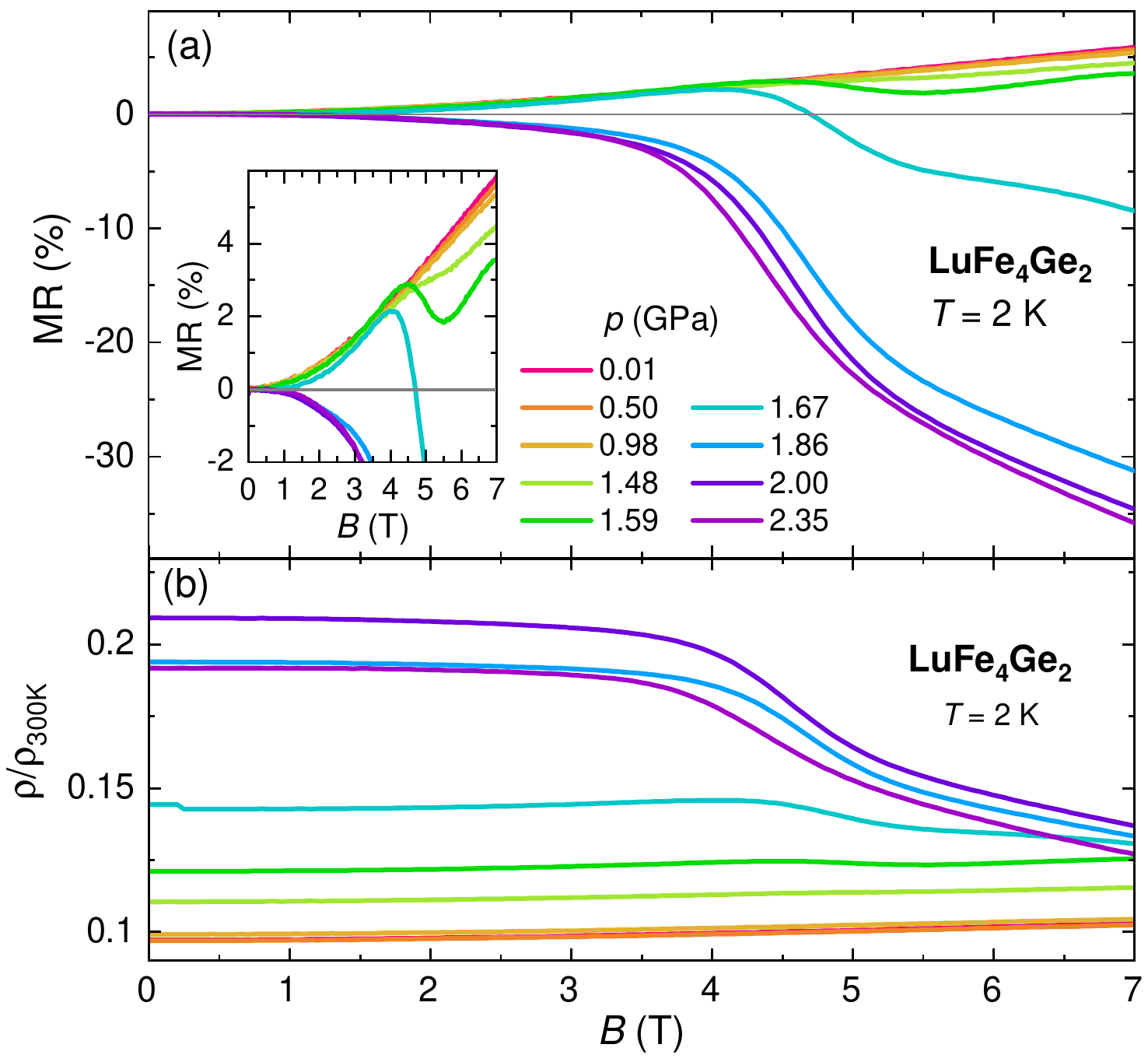}
\caption{(a) Magnetoresistance MR($B$) of LuFe$_4$Ge$_2$ measured at $T=2$~K for different pressures. The inset shows an enlarged view of the low-pressure curves. (b) Normalized resistivity $\rho/\rho_{\rm 300K}$ as a function of field at 2~K for several pressures.}
\label{Fig8}
\end{figure}

The transverse magnetoresistance MR$(B)=[\rho(B) -\rho(0)]/\rho(0)$ of LuFe$_4$Ge$_2$ as a function of magnetic field measured at different pressures at $T=2$~K is shown in Fig.~\ref{Fig8}a. MR$(B)$ presents evidence for a field-induced metamagnetic transition under pressure; a tiny kink in MR$(B)$ at about 5~T at pressures starting from 1.5~GPa (see inset of Fig.~\ref{Fig8}a). This feature becomes much pronounced upon increasing pressure and continuously shifts to lower fields. Eventually, a strong step-like decrease in the MR is observed at higher pressures with MR reaching $-35\%$ at 7~T for $p=2.35$~GPa. We note that high-field magnetization measurements on LuFe$_4$Ge$_2$ at ambient pressure revealed a weak metamagnetic transition at an applied magnetic field of 47~T, yet without the 
tendency of saturation of the magnetization~\cite{Weber17}. Such a metamagnetic transition could be related to a spin reorientation under the influence of applied magnetic field. This is also corroborated by the M$\ddot{\rm o}$ssbauer measurements at high pressure and a magnetic field of 6 T, where some part of the Fe moments seem to reorient in the direction of 
the applied field. It is also interesting to notice that the normalized resistivity $\rho/\rho_{\rm 300K}$ at higher magnetic fields appears to have similar values for all applied pressures, as displayed in Fig.~\ref{Fig8}b. This suggests that the field-polarized phase could be the same in the entire pressure range, once again indicating the close similarity between the AFM1 and AFM2 phases.

\newpage
\bibliography{LuFe4Ge2}

\providecommand{\noopsort}[1]{}\providecommand{\singleletter}[1]{#1}%
\begin{thebibliography}{26}%
\makeatletter
\providecommand \@ifxundefined [1]{%
 \@ifx{#1\undefined}
}%
\providecommand \@ifnum [1]{%
 \ifnum #1\expandafter \@firstoftwo
 \else \expandafter \@secondoftwo
 \fi
}%
\providecommand \@ifx [1]{%
 \ifx #1\expandafter \@firstoftwo
 \else \expandafter \@secondoftwo
 \fi
}%
\providecommand \natexlab [1]{#1}%
\providecommand \enquote  [1]{``#1''}%
\providecommand \bibnamefont  [1]{#1}%
\providecommand \bibfnamefont [1]{#1}%
\providecommand \citenamefont [1]{#1}%
\providecommand \href@noop [0]{\@secondoftwo}%
\providecommand \href [0]{\begingroup \@sanitize@url \@href}%
\providecommand \@href[1]{\@@startlink{#1}\@@href}%
\providecommand \@@href[1]{\endgroup#1\@@endlink}%
\providecommand \@sanitize@url [0]{\catcode `\\12\catcode `\$12\catcode
  `\&12\catcode `\#12\catcode `\^12\catcode `\_12\catcode `\%12\relax}%
\providecommand \@@startlink[1]{}%
\providecommand \@@endlink[0]{}%
\providecommand \url  [0]{\begingroup\@sanitize@url \@url }%
\providecommand \@url [1]{\endgroup\@href {#1}{\urlprefix }}%
\providecommand \urlprefix  [0]{URL }%
\providecommand \Eprint [0]{\href }%
\providecommand \doibase [0]{https://doi.org/}%
\providecommand \selectlanguage [0]{\@gobble}%
\providecommand \bibinfo  [0]{\@secondoftwo}%
\providecommand \bibfield  [0]{\@secondoftwo}%
\providecommand \translation [1]{[#1]}%
\providecommand \BibitemOpen [0]{}%
\providecommand \bibitemStop [0]{}%
\providecommand \bibitemNoStop [0]{.\EOS\space}%
\providecommand \EOS [0]{\spacefactor3000\relax}%
\providecommand \BibitemShut  [1]{\csname bibitem#1\endcsname}%
\let\auto@bib@innerbib\@empty
\bibitem [{\citenamefont {Mathur}\ \emph {et~al.}(1998)\citenamefont {Mathur},
  \citenamefont {Grosche}, \citenamefont {Julian}, \citenamefont {Walker},
  \citenamefont {Freye}, \citenamefont {Haselwimmer},\ and\ \citenamefont
  {Lonzarich}}]{Mathur98}%
  \BibitemOpen
  \bibfield  {author} {\bibinfo {author} {\bibfnamefont {N.~D.}\ \bibnamefont
  {Mathur}}, \bibinfo {author} {\bibfnamefont {F.~M.}\ \bibnamefont {Grosche}},
  \bibinfo {author} {\bibfnamefont {S.~R.}\ \bibnamefont {Julian}}, \bibinfo
  {author} {\bibfnamefont {I.~R.}\ \bibnamefont {Walker}}, \bibinfo {author}
  {\bibfnamefont {D.~M.}\ \bibnamefont {Freye}}, \bibinfo {author}
  {\bibfnamefont {R.~K.~W.}\ \bibnamefont {Haselwimmer}},\ and\ \bibinfo
  {author} {\bibfnamefont {G.~G.}\ \bibnamefont {Lonzarich}},\ }\bibfield
  {title} {\bibinfo {title} {Magnetically mediated superconductivity in heavy
  fermion compounds},\ }\href {https://doi.org/10.1038/27838} {\bibfield
  {journal} {\bibinfo  {journal} {Nature}\ }\textbf {\bibinfo {volume} {394}},\
  \bibinfo {pages} {39} (\bibinfo {year} {1998})}\BibitemShut {NoStop}%
\bibitem [{\citenamefont {Gegenwart}\ \emph {et~al.}(2008)\citenamefont
  {Gegenwart}, \citenamefont {Si},\ and\ \citenamefont
  {Steglich}}]{Gegenwart08}%
  \BibitemOpen
  \bibfield  {author} {\bibinfo {author} {\bibfnamefont {P.}~\bibnamefont
  {Gegenwart}}, \bibinfo {author} {\bibfnamefont {Q.}~\bibnamefont {Si}},\ and\
  \bibinfo {author} {\bibfnamefont {F.}~\bibnamefont {Steglich}},\ }\bibfield
  {title} {\bibinfo {title} {Quantum criticality in heavy-fermion metals},\
  }\href {https://doi.org/10.1038/nphys892} {\bibfield  {journal} {\bibinfo
  {journal} {Nat. Phys.}\ }\textbf {\bibinfo {volume} {4}},\ \bibinfo {pages}
  {186} (\bibinfo {year} {2008})}\BibitemShut {NoStop}%
\bibitem [{\citenamefont {Shibauchi}\ \emph {et~al.}(2014)\citenamefont
  {Shibauchi}, \citenamefont {Carrington},\ and\ \citenamefont
  {Matsuda}}]{Shibauchi14}%
  \BibitemOpen
  \bibfield  {author} {\bibinfo {author} {\bibfnamefont {T.}~\bibnamefont
  {Shibauchi}}, \bibinfo {author} {\bibfnamefont {A.}~\bibnamefont
  {Carrington}},\ and\ \bibinfo {author} {\bibfnamefont {Y.}~\bibnamefont
  {Matsuda}},\ }\bibfield  {title} {\bibinfo {title} {A quantum critical point
  lying beneath the superconducting dome in iron pnictides},\ }\href
  {https://doi.org/10.1146/annurev-conmatphys-031113-133921} {\bibfield
  {journal} {\bibinfo  {journal} {Annu. Rev. Condens. Matter Phys.}\ }\textbf
  {\bibinfo {volume} {5}},\ \bibinfo {pages} {113} (\bibinfo {year}
  {2014})}\BibitemShut {NoStop}%
\bibitem [{\citenamefont {Scalapino}(2012)}]{Scalapino12}%
  \BibitemOpen
  \bibfield  {author} {\bibinfo {author} {\bibfnamefont {D.~J.}\ \bibnamefont
  {Scalapino}},\ }\bibfield  {title} {\bibinfo {title} {A common thread: The
  pairing interaction for unconventional superconductors},\ }\href
  {https://doi.org/10.1103/RevModPhys.84.1383} {\bibfield  {journal} {\bibinfo
  {journal} {Rev. Mod. Phys.}\ }\textbf {\bibinfo {volume} {84}},\ \bibinfo
  {pages} {1383} (\bibinfo {year} {2012})}\BibitemShut {NoStop}%
\bibitem [{\citenamefont {Keimer}\ \emph {et~al.}(2015)\citenamefont {Keimer},
  \citenamefont {Kivelson}, \citenamefont {Norman}, \citenamefont {Uchida},\
  and\ \citenamefont {Zaanen}}]{Keimer15}%
  \BibitemOpen
  \bibfield  {author} {\bibinfo {author} {\bibfnamefont {B.}~\bibnamefont
  {Keimer}}, \bibinfo {author} {\bibfnamefont {S.}~\bibnamefont {Kivelson}},
  \bibinfo {author} {\bibfnamefont {M.}~\bibnamefont {Norman}}, \bibinfo
  {author} {\bibfnamefont {S.}~\bibnamefont {Uchida}},\ and\ \bibinfo {author}
  {\bibfnamefont {J.}~\bibnamefont {Zaanen}},\ }\bibfield  {title} {\bibinfo
  {title} {From quantum matter to high-temperature superconductivity in copper
  oxides},\ }\href {https://doi.org/10.1038/nature14165} {\bibfield  {journal}
  {\bibinfo  {journal} {Nature}\ }\textbf {\bibinfo {volume} {518}},\ \bibinfo
  {pages} {179} (\bibinfo {year} {2015})}\BibitemShut {NoStop}%
\bibitem [{\citenamefont {Gruner}\ \emph {et~al.}(2017)\citenamefont {Gruner},
  \citenamefont {Jang}, \citenamefont {Huesges}, \citenamefont {Cardoso-Gil},
  \citenamefont {Fecher}, \citenamefont {Koza}, \citenamefont {Stockert},
  \citenamefont {Mackenzie}, \citenamefont {Brando},\ and\ \citenamefont
  {Geibel}}]{Gruner17}%
  \BibitemOpen
  \bibfield  {author} {\bibinfo {author} {\bibfnamefont {T.}~\bibnamefont
  {Gruner}}, \bibinfo {author} {\bibfnamefont {D.}~\bibnamefont {Jang}},
  \bibinfo {author} {\bibfnamefont {Z.}~\bibnamefont {Huesges}}, \bibinfo
  {author} {\bibfnamefont {R.}~\bibnamefont {Cardoso-Gil}}, \bibinfo {author}
  {\bibfnamefont {G.~H.}\ \bibnamefont {Fecher}}, \bibinfo {author}
  {\bibfnamefont {M.~M.}\ \bibnamefont {Koza}}, \bibinfo {author}
  {\bibfnamefont {O.}~\bibnamefont {Stockert}}, \bibinfo {author}
  {\bibfnamefont {A.~P.}\ \bibnamefont {Mackenzie}}, \bibinfo {author}
  {\bibfnamefont {M.}~\bibnamefont {Brando}},\ and\ \bibinfo {author}
  {\bibfnamefont {C.}~\bibnamefont {Geibel}},\ }\bibfield  {title} {\bibinfo
  {title} {Charge density wave quantum critical point with strong enhancement
  of superconductivity},\ }\href {https://doi.org/10.1038/nphys4191} {\bibfield
   {journal} {\bibinfo  {journal} {Nat. Phys.}\ }\textbf {\bibinfo {volume}
  {13}},\ \bibinfo {pages} {967} (\bibinfo {year} {2017})}\BibitemShut
  {NoStop}%
\bibitem [{\citenamefont {Chubukov}(2012)}]{Chubukov12}%
  \BibitemOpen
  \bibfield  {author} {\bibinfo {author} {\bibfnamefont {A.}~\bibnamefont
  {Chubukov}},\ }\bibfield  {title} {\bibinfo {title} {Pairing mechanism in
  {Fe}-based superconductors},\ }\href
  {https://doi.org/10.1146/annurev-conmatphys-020911-125055} {\bibfield
  {journal} {\bibinfo  {journal} {Annu. Rev. Condens. Matter Phys.}\ }\textbf
  {\bibinfo {volume} {3}},\ \bibinfo {pages} {57} (\bibinfo {year}
  {2012})}\BibitemShut {NoStop}%
\bibitem [{\citenamefont {Dai}(2015)}]{Dai15}%
  \BibitemOpen
  \bibfield  {author} {\bibinfo {author} {\bibfnamefont {P.}~\bibnamefont
  {Dai}},\ }\bibfield  {title} {\bibinfo {title} {Antiferromagnetic order and
  spin dynamics in iron-based superconductors},\ }\href
  {https://doi.org/10.1103/RevModPhys.87.855} {\bibfield  {journal} {\bibinfo
  {journal} {Rev. Mod. Phys.}\ }\textbf {\bibinfo {volume} {87}},\ \bibinfo
  {pages} {855} (\bibinfo {year} {2015})}\BibitemShut {NoStop}%
\bibitem [{\citenamefont {Inosov}(2016)}]{Inosov16}%
  \BibitemOpen
  \bibfield  {author} {\bibinfo {author} {\bibfnamefont {D.~S.}\ \bibnamefont
  {Inosov}},\ }\bibfield  {title} {\bibinfo {title} {Spin fluctuations in iron
  pnictides and chalcogenides: From antiferromagnetism to superconductivity},\
  }\href {https://doi.org/https://doi.org/10.1016/j.crhy.2015.03.001}
  {\bibfield  {journal} {\bibinfo  {journal} {C. R. Phys.}\ }\textbf {\bibinfo
  {volume} {17}},\ \bibinfo {pages} {60} (\bibinfo {year} {2016})}\BibitemShut
  {NoStop}%
\bibitem [{\citenamefont {Kuo}\ \emph {et~al.}(2016)\citenamefont {Kuo},
  \citenamefont {Chu}, \citenamefont {Palmstrom}, \citenamefont {Kivelson},\
  and\ \citenamefont {Fisher}}]{Kou16}%
  \BibitemOpen
  \bibfield  {author} {\bibinfo {author} {\bibfnamefont {H.-H.}\ \bibnamefont
  {Kuo}}, \bibinfo {author} {\bibfnamefont {J.-H.}\ \bibnamefont {Chu}},
  \bibinfo {author} {\bibfnamefont {J.~C.}\ \bibnamefont {Palmstrom}}, \bibinfo
  {author} {\bibfnamefont {S.~A.}\ \bibnamefont {Kivelson}},\ and\ \bibinfo
  {author} {\bibfnamefont {I.~R.}\ \bibnamefont {Fisher}},\ }\bibfield  {title}
  {\bibinfo {title} {Ubiquitous signatures of nematic quantum criticality in
  optimally doped {Fe}-based superconductors},\ }\href
  {https://doi.org/10.1126/science.aab0103} {\bibfield  {journal} {\bibinfo
  {journal} {Science}\ }\textbf {\bibinfo {volume} {352}},\ \bibinfo {pages}
  {958} (\bibinfo {year} {2016})}\BibitemShut {NoStop}%
\bibitem [{\citenamefont {Wang}\ \emph {et~al.}(2022)\citenamefont {Wang},
  \citenamefont {Fanfarillo},\ and\ \citenamefont {Böhmer}}]{Wang22}%
  \BibitemOpen
  \bibfield  {author} {\bibinfo {author} {\bibfnamefont {Q.}~\bibnamefont
  {Wang}}, \bibinfo {author} {\bibfnamefont {L.}~\bibnamefont {Fanfarillo}},\
  and\ \bibinfo {author} {\bibfnamefont {A.~E.}\ \bibnamefont {Böhmer}},\
  }\bibfield  {title} {\bibinfo {title} {Editorial: Nematicity in iron-based
  superconductors},\ }\href {https://doi.org/10.3389/fphy.2022.1038127}
  {\bibfield  {journal} {\bibinfo  {journal} {Front. Phys.}\ }\textbf {\bibinfo
  {volume} {10}},\ \bibinfo {pages} {1038127} (\bibinfo {year}
  {2022})}\BibitemShut {NoStop}%
\bibitem [{\citenamefont {Yarmoluk}\ \emph {et~al.}(1975)\citenamefont
  {Yarmoluk}, \citenamefont {Lysenko},\ and\ \citenamefont
  {Gladyshevski}}]{Yarmoluk75}%
  \BibitemOpen
  \bibfield  {author} {\bibinfo {author} {\bibfnamefont {Y.~P.}\ \bibnamefont
  {Yarmoluk}}, \bibinfo {author} {\bibfnamefont {L.~A.}\ \bibnamefont
  {Lysenko}},\ and\ \bibinfo {author} {\bibfnamefont {E.~I.}\ \bibnamefont
  {Gladyshevski}},\ }\bibfield  {title} {\bibinfo {title} {Zirconium iron
  silicide (\text{ZrFe$_4$Si$_2$}) structure as a new structural type of
  ternary transition metal silicides},\ }\href@noop {} {\bibfield  {journal}
  {\bibinfo  {journal} {Dopov. Nac. akad. nauk Ukr. RSR, Series A.}\ }\textbf
  {\bibinfo {volume} {37}},\ \bibinfo {pages} {279} (\bibinfo {year}
  {1975})}\BibitemShut {NoStop}%
\bibitem [{\citenamefont {Ajeesh}\ \emph {et~al.}(2020)\citenamefont {Ajeesh},
  \citenamefont {Weber}, \citenamefont {Geibel},\ and\ \citenamefont
  {Nicklas}}]{Ajeesh20}%
  \BibitemOpen
  \bibfield  {author} {\bibinfo {author} {\bibfnamefont {M.~O.}\ \bibnamefont
  {Ajeesh}}, \bibinfo {author} {\bibfnamefont {K.}~\bibnamefont {Weber}},
  \bibinfo {author} {\bibfnamefont {C.}~\bibnamefont {Geibel}},\ and\ \bibinfo
  {author} {\bibfnamefont {M.}~\bibnamefont {Nicklas}},\ }\bibfield  {title}
  {\bibinfo {title} {Putative quantum critical point in the itinerant magnet
  \text{${\mathrm{ZrFe}}_{4}{\mathrm{Si}}_{2}$} with a frustrated
  quasi-one-dimensional structure},\ }\href
  {https://doi.org/10.1103/PhysRevB.102.184403} {\bibfield  {journal} {\bibinfo
   {journal} {Phys. Rev. B}\ }\textbf {\bibinfo {volume} {102}},\ \bibinfo
  {pages} {184403} (\bibinfo {year} {2020})}\BibitemShut {NoStop}%
\bibitem [{\citenamefont {Schobinger-Papamantellos}\ \emph
  {et~al.}(2012)\citenamefont {Schobinger-Papamantellos}, \citenamefont
  {Buschow},\ and\ \citenamefont {Rodríguez-Carvajal}}]{Schobinger12}%
  \BibitemOpen
  \bibfield  {author} {\bibinfo {author} {\bibfnamefont {P.}~\bibnamefont
  {Schobinger-Papamantellos}}, \bibinfo {author} {\bibfnamefont {K.~H.~J.}\
  \bibnamefont {Buschow}},\ and\ \bibinfo {author} {\bibfnamefont
  {J.}~\bibnamefont {Rodríguez-Carvajal}},\ }\bibfield  {title} {\bibinfo
  {title} {Magnetoelastic phase transitions in the \text{LuFe$_4$Ge$_2$} and
  \text{YFe$_4$Si$_2$} compounds: A neutron diffraction study},\ }\href
  {https://doi.org/10.1016/j.jmmm.2012.05.058} {\bibfield  {journal} {\bibinfo
  {journal} {J. Magn. Magn. Mater.}\ }\textbf {\bibinfo {volume} {324}},\
  \bibinfo {pages} {3709} (\bibinfo {year} {2012})}\BibitemShut {NoStop}%
\bibitem [{\citenamefont {Lashley}\ \emph {et~al.}(2003)\citenamefont
  {Lashley}, \citenamefont {Hundley}, \citenamefont {Migliori}, \citenamefont
  {Sarrao}, \citenamefont {Pagliuso}, \citenamefont {Darling}, \citenamefont
  {Jaime}, \citenamefont {Cooley}, \citenamefont {Hults}, \citenamefont
  {Morales}, \citenamefont {Thoma}, \citenamefont {Smith}, \citenamefont
  {Boerio-Goates}, \citenamefont {Woodfield}, \citenamefont {Stewart},
  \citenamefont {Fisher},\ and\ \citenamefont {Phillips}}]{Lashley03}%
  \BibitemOpen
  \bibfield  {author} {\bibinfo {author} {\bibfnamefont {J.}~\bibnamefont
  {Lashley}}, \bibinfo {author} {\bibfnamefont {M.}~\bibnamefont {Hundley}},
  \bibinfo {author} {\bibfnamefont {A.}~\bibnamefont {Migliori}}, \bibinfo
  {author} {\bibfnamefont {J.}~\bibnamefont {Sarrao}}, \bibinfo {author}
  {\bibfnamefont {P.}~\bibnamefont {Pagliuso}}, \bibinfo {author}
  {\bibfnamefont {T.}~\bibnamefont {Darling}}, \bibinfo {author} {\bibfnamefont
  {M.}~\bibnamefont {Jaime}}, \bibinfo {author} {\bibfnamefont
  {J.}~\bibnamefont {Cooley}}, \bibinfo {author} {\bibfnamefont
  {W.}~\bibnamefont {Hults}}, \bibinfo {author} {\bibfnamefont
  {L.}~\bibnamefont {Morales}}, \bibinfo {author} {\bibfnamefont
  {D.}~\bibnamefont {Thoma}}, \bibinfo {author} {\bibfnamefont
  {J.}~\bibnamefont {Smith}}, \bibinfo {author} {\bibfnamefont
  {J.}~\bibnamefont {Boerio-Goates}}, \bibinfo {author} {\bibfnamefont
  {B.}~\bibnamefont {Woodfield}}, \bibinfo {author} {\bibfnamefont
  {G.}~\bibnamefont {Stewart}}, \bibinfo {author} {\bibfnamefont
  {R.}~\bibnamefont {Fisher}},\ and\ \bibinfo {author} {\bibfnamefont
  {N.}~\bibnamefont {Phillips}},\ }\bibfield  {title} {\bibinfo {title}
  {Critical examination of heat capacity measurements made on a {Quantum
  Design} physical property measurement system},\ }\href
  {https://doi.org/https://doi.org/10.1016/S0011-2275(03)00092-4} {\bibfield
  {journal} {\bibinfo  {journal} {Cryogenics}\ }\textbf {\bibinfo {volume}
  {43}},\ \bibinfo {pages} {369} (\bibinfo {year} {2003})}\BibitemShut
  {NoStop}%
\bibitem [{\citenamefont {Weber}(2017)}]{Weber17}%
  \BibitemOpen
  \bibfield  {author} {\bibinfo {author} {\bibfnamefont {K.}~\bibnamefont
  {Weber}},\ }\bibfield  {title} {\bibinfo {title} {Phd thesis},\ }\href@noop
  {} {\bibfield  {journal} {\bibinfo  {journal} {Technical University of
  Dresden, Dresden, Germany}\ } (\bibinfo {year} {2017})}\BibitemShut {NoStop}%
\bibitem [{\citenamefont {Gati}\ \emph {et~al.}(2019)\citenamefont {Gati},
  \citenamefont {Drachuck}, \citenamefont {Xiang}, \citenamefont {Wang},
  \citenamefont {Bud’ko},\ and\ \citenamefont {Canfield}}]{Gati19}%
  \BibitemOpen
  \bibfield  {author} {\bibinfo {author} {\bibfnamefont {E.}~\bibnamefont
  {Gati}}, \bibinfo {author} {\bibfnamefont {G.}~\bibnamefont {Drachuck}},
  \bibinfo {author} {\bibfnamefont {L.}~\bibnamefont {Xiang}}, \bibinfo
  {author} {\bibfnamefont {L.-L.}\ \bibnamefont {Wang}}, \bibinfo {author}
  {\bibfnamefont {S.~L.}\ \bibnamefont {Bud’ko}},\ and\ \bibinfo {author}
  {\bibfnamefont {P.~C.}\ \bibnamefont {Canfield}},\ }\bibfield  {title}
  {\bibinfo {title} {Use of \text{Cernox} thermometers in \text{AC} specific
  heat measurements under pressure},\ }\href
  {https://doi.org/10.1063/1.5084730} {\bibfield  {journal} {\bibinfo
  {journal} {Rev. Sci. Instrum.}\ }\textbf {\bibinfo {volume} {90}},\ \bibinfo
  {pages} {023911} (\bibinfo {year} {2019})}\BibitemShut {NoStop}%
\bibitem [{\citenamefont {Khasanov}\ \emph {et~al.}(2016)\citenamefont
  {Khasanov}, \citenamefont {Guguchia}, \citenamefont {Maisuradze},
  \citenamefont {Andreica}, \citenamefont {Elender}, \citenamefont {Raselli},
  \citenamefont {Shermadini}, \citenamefont {Goko}, \citenamefont {Knecht},
  \citenamefont {Morenzoni},\ and\ \citenamefont {Amato}}]{Khasanov16}%
  \BibitemOpen
  \bibfield  {author} {\bibinfo {author} {\bibfnamefont {R.}~\bibnamefont
  {Khasanov}}, \bibinfo {author} {\bibfnamefont {Z.}~\bibnamefont {Guguchia}},
  \bibinfo {author} {\bibfnamefont {A.}~\bibnamefont {Maisuradze}}, \bibinfo
  {author} {\bibfnamefont {D.}~\bibnamefont {Andreica}}, \bibinfo {author}
  {\bibfnamefont {M.}~\bibnamefont {Elender}}, \bibinfo {author} {\bibfnamefont
  {A.}~\bibnamefont {Raselli}}, \bibinfo {author} {\bibfnamefont
  {Z.}~\bibnamefont {Shermadini}}, \bibinfo {author} {\bibfnamefont
  {T.}~\bibnamefont {Goko}}, \bibinfo {author} {\bibfnamefont {F.}~\bibnamefont
  {Knecht}}, \bibinfo {author} {\bibfnamefont {E.}~\bibnamefont {Morenzoni}},\
  and\ \bibinfo {author} {\bibfnamefont {A.}~\bibnamefont {Amato}},\ }\bibfield
   {title} {\bibinfo {title} {High pressure research using muons at the
  \text{Paul Scherrer Institute}},\ }\href
  {https://doi.org/10.1080/08957959.2016.1173690} {\bibfield  {journal}
  {\bibinfo  {journal} {High Press. Res.}\ }\textbf {\bibinfo {volume} {36}},\
  \bibinfo {pages} {140} (\bibinfo {year} {2016})}\BibitemShut {NoStop}%
\bibitem [{\citenamefont {Khasanov}(2022)}]{Khasanov22}%
  \BibitemOpen
  \bibfield  {author} {\bibinfo {author} {\bibfnamefont {R.}~\bibnamefont
  {Khasanov}},\ }\bibfield  {title} {\bibinfo {title} {Perspective on muon-spin
  rotation/relaxation under hydrostatic pressure},\ }\href
  {https://doi.org/10.1063/5.0119840} {\bibfield  {journal} {\bibinfo
  {journal} {Journal of Applied Physics}\ }\textbf {\bibinfo {volume} {132}},\
  \bibinfo {pages} {190903} (\bibinfo {year} {2022})}\BibitemShut {NoStop}%
\bibitem [{\citenamefont {Potapkin}\ \emph {et~al.}(2012)\citenamefont
  {Potapkin}, \citenamefont {Chumakov}, \citenamefont {Smirnov}, \citenamefont
  {Celse}, \citenamefont {R{\"{u}}ffer}, \citenamefont {McCammon},\ and\
  \citenamefont {Dubrovinsky}}]{Potapkin12}%
  \BibitemOpen
  \bibfield  {author} {\bibinfo {author} {\bibfnamefont {V.}~\bibnamefont
  {Potapkin}}, \bibinfo {author} {\bibfnamefont {A.~I.}\ \bibnamefont
  {Chumakov}}, \bibinfo {author} {\bibfnamefont {G.~V.}\ \bibnamefont
  {Smirnov}}, \bibinfo {author} {\bibfnamefont {J.-P.}\ \bibnamefont {Celse}},
  \bibinfo {author} {\bibfnamefont {R.}~\bibnamefont {R{\"{u}}ffer}}, \bibinfo
  {author} {\bibfnamefont {C.}~\bibnamefont {McCammon}},\ and\ \bibinfo
  {author} {\bibfnamefont {L.}~\bibnamefont {Dubrovinsky}},\ }\bibfield
  {title} {\bibinfo {title} {{The $^{57}${Fe} Synchrotron {M{\"{o}}ssbauer}
  Source at the \text{ESRF}}},\ }\href
  {https://doi.org/10.1107/S0909049512015579} {\bibfield  {journal} {\bibinfo
  {journal} {J. of Synchrotron Radiat.}\ }\textbf {\bibinfo {volume} {19}},\
  \bibinfo {pages} {559} (\bibinfo {year} {2012})}\BibitemShut {NoStop}%
\bibitem [{\citenamefont {Bi}\ \emph {et~al.}(2015)\citenamefont {Bi},
  \citenamefont {Zhao}, \citenamefont {Lin}, \citenamefont {Jia}, \citenamefont
  {Hu}, \citenamefont {Jin}, \citenamefont {Ferry}, \citenamefont {Yang},
  \citenamefont {Struzhkin},\ and\ \citenamefont {Alp}}]{Bi15}%
  \BibitemOpen
  \bibfield  {author} {\bibinfo {author} {\bibfnamefont {W.}~\bibnamefont
  {Bi}}, \bibinfo {author} {\bibfnamefont {J.}~\bibnamefont {Zhao}}, \bibinfo
  {author} {\bibfnamefont {J.-F.}\ \bibnamefont {Lin}}, \bibinfo {author}
  {\bibfnamefont {Q.}~\bibnamefont {Jia}}, \bibinfo {author} {\bibfnamefont
  {M.~Y.}\ \bibnamefont {Hu}}, \bibinfo {author} {\bibfnamefont
  {C.}~\bibnamefont {Jin}}, \bibinfo {author} {\bibfnamefont {R.}~\bibnamefont
  {Ferry}}, \bibinfo {author} {\bibfnamefont {W.}~\bibnamefont {Yang}},
  \bibinfo {author} {\bibfnamefont {V.}~\bibnamefont {Struzhkin}},\ and\
  \bibinfo {author} {\bibfnamefont {E.~E.}\ \bibnamefont {Alp}},\ }\bibfield
  {title} {\bibinfo {title} {{Nuclear resonant inelastic X-ray scattering at
  high pressure and low temperature}},\ }\href
  {https://doi.org/10.1107/S1600577515003586} {\bibfield  {journal} {\bibinfo
  {journal} {J. Synchrotron Radiat.}\ }\textbf {\bibinfo {volume} {22}},\
  \bibinfo {pages} {760} (\bibinfo {year} {2015})}\BibitemShut {NoStop}%
\bibitem [{\citenamefont {Zhao}\ \emph {et~al.}(2017)\citenamefont {Zhao},
  \citenamefont {Bi}, \citenamefont {Sinogeikin}, \citenamefont {Hu},
  \citenamefont {Alp}, \citenamefont {Wang}, \citenamefont {Jin},\ and\
  \citenamefont {Lin}}]{Zhao17}%
  \BibitemOpen
  \bibfield  {author} {\bibinfo {author} {\bibfnamefont {J.~Y.}\ \bibnamefont
  {Zhao}}, \bibinfo {author} {\bibfnamefont {W.}~\bibnamefont {Bi}}, \bibinfo
  {author} {\bibfnamefont {S.}~\bibnamefont {Sinogeikin}}, \bibinfo {author}
  {\bibfnamefont {M.~Y.}\ \bibnamefont {Hu}}, \bibinfo {author} {\bibfnamefont
  {E.~E.}\ \bibnamefont {Alp}}, \bibinfo {author} {\bibfnamefont {X.~C.}\
  \bibnamefont {Wang}}, \bibinfo {author} {\bibfnamefont {C.~Q.}\ \bibnamefont
  {Jin}},\ and\ \bibinfo {author} {\bibfnamefont {J.~F.}\ \bibnamefont {Lin}},\
  }\bibfield  {title} {\bibinfo {title} {A compact membrane-driven diamond
  anvil cell and cryostat system for nuclear resonant scattering at high
  pressure and low temperature},\ }\href {https://doi.org/10.1063/1.4999787}
  {\bibfield  {journal} {\bibinfo  {journal} {Rev. Sci. Instrum.}\ }\textbf
  {\bibinfo {volume} {88}},\ \bibinfo {pages} {125109} (\bibinfo {year}
  {2017})}\BibitemShut {NoStop}%
\bibitem [{\citenamefont {Lima}\ \emph {et~al.}(2016)\citenamefont {Lima},
  \citenamefont {Saleta}, \citenamefont {Pagliuca}, \citenamefont
  {Eleot{\'{e}}rio}, \citenamefont {Reis}, \citenamefont
  {Fonseca~J{\'{u}}nior}, \citenamefont {Meyer}, \citenamefont {Bittar},
  \citenamefont {Souza-Neto},\ and\ \citenamefont {Granado}}]{Lima16}%
  \BibitemOpen
  \bibfield  {author} {\bibinfo {author} {\bibfnamefont {F.~A.}\ \bibnamefont
  {Lima}}, \bibinfo {author} {\bibfnamefont {M.~E.}\ \bibnamefont {Saleta}},
  \bibinfo {author} {\bibfnamefont {R.~J.~S.}\ \bibnamefont {Pagliuca}},
  \bibinfo {author} {\bibfnamefont {M.~A.}\ \bibnamefont {Eleot{\'{e}}rio}},
  \bibinfo {author} {\bibfnamefont {R.~D.}\ \bibnamefont {Reis}}, \bibinfo
  {author} {\bibfnamefont {J.}~\bibnamefont {Fonseca~J{\'{u}}nior}}, \bibinfo
  {author} {\bibfnamefont {B.}~\bibnamefont {Meyer}}, \bibinfo {author}
  {\bibfnamefont {E.~M.}\ \bibnamefont {Bittar}}, \bibinfo {author}
  {\bibfnamefont {N.~M.}\ \bibnamefont {Souza-Neto}},\ and\ \bibinfo {author}
  {\bibfnamefont {E.}~\bibnamefont {Granado}},\ }\bibfield  {title} {\bibinfo
  {title} {{XDS: a flexible beamline for X-ray diffraction and spectroscopy at
  the Brazilian synchrotron}},\ }\href
  {https://doi.org/10.1107/S160057751601403X} {\bibfield  {journal} {\bibinfo
  {journal} {J. Synchrotron Radiat.}\ }\textbf {\bibinfo {volume} {23}},\
  \bibinfo {pages} {1538} (\bibinfo {year} {2016})}\BibitemShut {NoStop}%
\bibitem [{\citenamefont {Kresse}\ and\ \citenamefont
  {Furthm\"uller}(1996)}]{VASP}%
  \BibitemOpen
  \bibfield  {author} {\bibinfo {author} {\bibfnamefont {G.}~\bibnamefont
  {Kresse}}\ and\ \bibinfo {author} {\bibfnamefont {J.}~\bibnamefont
  {Furthm\"uller}},\ }\bibfield  {title} {\bibinfo {title} {Efficient iterative
  schemes for ab initio total-energy calculations using a plane-wave basis
  set},\ }\href {https://doi.org/10.1103/PhysRevB.54.11169} {\bibfield
  {journal} {\bibinfo  {journal} {Phys. Rev. B}\ }\textbf {\bibinfo {volume}
  {54}},\ \bibinfo {pages} {11169} (\bibinfo {year} {1996})}\BibitemShut
  {NoStop}%
\bibitem [{\citenamefont {Perdew}\ and\ \citenamefont {Wang}(1992)}]{PW}%
  \BibitemOpen
  \bibfield  {author} {\bibinfo {author} {\bibfnamefont {J.~P.}\ \bibnamefont
  {Perdew}}\ and\ \bibinfo {author} {\bibfnamefont {Y.}~\bibnamefont {Wang}},\
  }\bibfield  {title} {\bibinfo {title} {Accurate and simple analytic
  representation of the electron-gas correlation energy},\ }\href
  {https://doi.org/10.1103/PhysRevB.45.13244} {\bibfield  {journal} {\bibinfo
  {journal} {Phys. Rev. B}\ }\textbf {\bibinfo {volume} {45}},\ \bibinfo
  {pages} {13244} (\bibinfo {year} {1992})}\BibitemShut {NoStop}%
\bibitem [{\citenamefont {Perdew}\ \emph {et~al.}(1996)\citenamefont {Perdew},
  \citenamefont {Burke},\ and\ \citenamefont {Ernzerhof}}]{PBE}%
  \BibitemOpen
  \bibfield  {author} {\bibinfo {author} {\bibfnamefont {J.~P.}\ \bibnamefont
  {Perdew}}, \bibinfo {author} {\bibfnamefont {K.}~\bibnamefont {Burke}},\ and\
  \bibinfo {author} {\bibfnamefont {M.}~\bibnamefont {Ernzerhof}},\ }\bibfield
  {title} {\bibinfo {title} {Generalized gradient approximation made simple},\
  }\href {https://doi.org/10.1103/PhysRevLett.77.3865} {\bibfield  {journal}
  {\bibinfo  {journal} {Phys. Rev. Lett.}\ }\textbf {\bibinfo {volume} {77}},\
  \bibinfo {pages} {3865} (\bibinfo {year} {1996})}\BibitemShut {NoStop}%
\end{thebibliography}%

\end{document}